\documentclass[journal, a4paper]{IEEEtran}
\usepackage{algorithmicx}
\usepackage{algorithm}
\usepackage{subcaption}
\usepackage{float}
\usepackage{bm}
\usepackage{upgreek}
\usepackage{diagbox}
\usepackage{amsthm}

\usepackage{tabularx}
\usepackage{comment}
\usepackage{algorithm}
\usepackage{makecell}
\usepackage{titlesec}
\usepackage{indentfirst}
\usepackage{algpseudocode}
\usepackage{booktabs,multirow,tabularx}
\usepackage{bbm}
\usepackage{ulem}

\makeatletter
\newcommand{\removelatexerror}{\let\@latex@error\@gobble}
\makeatother
\IEEEoverridecommandlockouts

\usepackage{cite}
\usepackage{xcolor}

\usepackage[colorlinks,
hyperfootnotes=false,
linkcolor=blue,
anchorcolor=blue,
citecolor=blue,
urlcolor=black]{hyperref}
\usepackage{amsmath,amssymb,amsfonts}
\allowdisplaybreaks
\usepackage{mathtools}
\usepackage{graphicx}
\usepackage{textcomp}
\def\BibTeX{{\rm B\kern-.05em{\sc i\kern-.025em b}\kern-.08em
        T\kern-.1667em\lower.7ex\hbox{E}\kern-.125emX}}

\usepackage{setspace}

\begin{document}
\flushbottom
\bstctlcite{IEEEexample:BSTcontrol}
\algnewcommand\algorithmicswitch{\textbf{switch}}
\algnewcommand\algorithmiccase{\textbf{case}}
\algnewcommand\algorithmicassert{\texttt{assert}}
\algnewcommand\Assert[1]{\State \algorithmicassert(#1)}%

\algdef{SE}[SWITCH]{Switch}{EndSwitch}[1]{\algorithmicswitch\ #1\ \algorithmicdo}{\algorithmicend\ \algorithmicswitch}%
\algdef{SE}[CASE]{Case}{EndCase}[1]{\algorithmiccase\ #1}{\algorithmicend\ \algorithmiccase}%
\algtext*{EndSwitch}%
\algtext*{EndCase}%

\title{Multi-Modal Environment-Aware \\Beam Management for Massive MIMO: \\A Geometry-Driven Virtual Base Station Framework}
\author{Yijie~Bian,~Wei~Guo,~\IEEEmembership{Member,~IEEE},~Jie~Yang,~\IEEEmembership{Member,~IEEE},~Shenghui~Song,~\IEEEmembership{Senior~Member,~IEEE},\\~Jun~Zhang,~\IEEEmembership{Fellow,~IEEE},~Shi~Jin,~\IEEEmembership{Fellow,~IEEE},~Khaled~B.~Letaief,~\IEEEmembership{Fellow,~IEEE}
\vspace{-1. cm}
\thanks{
An earlier version of this paper was presented in part at the IEEE Wireless Communications and Networking Conference (WCNC), Kuala Lumpur, Malaysia, in April 2026 \cite{vbs_beam_alignment}.

Y. Bian, W. Guo, S. Song, J. Zhang, and K. B. Letaief are with the Department of Electronic and Computer Engineering, Hong Kong University of Science and Technology, Hong Kong SAR. Emails: ybianaf@connect.ust.hk, \{eeweiguo, eeshsong, eejzhang, eekhaled\}@ust.hk.

J. Yang is with the School of Automation, Southeast University, Nanjing, China. Email: yangjie@seu.edu.cn.

S. Jin is with the School of Information Science and Engineering, Southeast University, Nanjing, China. Email: jinshi@seu.edu.cn.
}
}

\maketitle
\begin{abstract}
High-frequency massive multiple-input multiple-output (MIMO) systems promise ultra-high data rates. However, efficient beam management remains challenging due to the prohibitive beam training overhead and intricate coordination required in multi-user MIMO (MU-MIMO) scenarios. To address these bottlenecks, environment-aware communications have emerged as a promising paradigm, leveraging site-specific knowledge to circumvent exhaustive pilot-based beam training and streamline multi-user communications. In this paper, we propose a geometry-driven framework that utilizes multi-modal environmental data, specifically regional 3D light detection and ranging (LiDAR) point clouds and location information, to construct an offline virtual base station (VBS) database. By modeling dominant reflection paths via mirror symmetry across building facades reconstructed from the point clouds, the VBS database provides a compact and sparse description of the wireless propagation environment. Given a user location, the VBS representation supports online 
coarse channel reconstruction and the generation of
a beamspace prior. To 
refine the coarse estimates, we design a VBS-assisted orthogonal-pilot (VOP)-based partial
beam training scheme with limited
online measurements. Finally, to coordinate beam selection under inter-user interference,
we develop a hierarchical dual-agent dueling double
deep Q-network for coordinated beam selection (DD3QN-CBS).  Simulation results show that the proposed framework improves effective spectral efficiency over the evaluated stochastic, heuristic, and conventional learning-based baselines, while achieving competitive performance with fewer online beam measurements.
\end{abstract}
\begin{IEEEkeywords}
Virtual base station (VBS), multi-user multiple-input multiple-output (MU-MIMO) beam management, multi-modal data, deep reinforcement learning (DRL)
\end{IEEEkeywords}

\section{Introduction}
\label{Introduction}

To meet the exponential growth in data traffic and throughput requirements of the sixth-generation (6G) networks, exploring spectrum resources at increasingly elevated frequencies has become imperative \cite{roadmap_6g}. High-frequency bands, ranging from millimeter-wave (mmWave) to sub-terahertz (sub-THz) and terahertz (THz), are poised to serve as cornerstones of future wireless systems. These bands offer ultra-high bandwidths capable of supporting massive data throughput and enhanced connectivity, thereby enabling advanced applications in both outdoor and indoor scenarios, such as holographic telepresence and immersive extended reality \cite{roadmap_6g}. However, these high-frequency bands suffer from high blockage sensitivity, low diffraction capability, and increased propagation attenuation. To compensate for this attenuation, massive multiple-input multiple-output (MIMO) technology employing large-scale antenna arrays has been widely adopted. By leveraging high array gains through beamforming and spatial multiplexing, massive MIMO systems can significantly enhance link reliability and spectral efficiency \cite{hybrid_beamforming,hybrid_mmWave_MMSE}.

Nevertheless, realizing the full potential of massive MIMO in high-frequency bands presents significant challenges. Due to hardware cost and power consumption constraints, transceivers typically adopt analog or hybrid architectures in which the number of radio frequency (RF) chains is much smaller than that of antenna elements. Such hardware constraints restrict the system to low-dimensional observations, making it computationally intensive and time-prohibitive to acquire accurate high-dimensional channel state information (CSI) in real time. Consequently, practical standards and existing studies resort to codebook-based beam management \cite{tutorial_bm,survey_beam_management_6g}. Instead of conducting explicit channel estimation, the base station (BS) and user select the optimal beam pair from predefined codebooks through beam training. While feasible, this approach introduces a critical bottleneck that exhaustive beam training incurs substantial overhead and latency \cite{meta_beam_alignment}. The challenge becomes even more pronounced in multi-user scenarios, where the network must coordinate beam selection across users to mitigate inter-user interference, leading to a combinatorial search over beam pairs \cite{joint_beam_selection_precoding}.

\subsection{Prior Works and Limitations}

To address the prohibitively high overhead of acquiring real-time CSI for beam management, a promising strategy is to exploit environmental sensing data, such as light detection and ranging (LiDAR) point clouds \cite{lidar_beam_prediction,lidar_beam_selection}, radio detection and ranging (Radar) images \cite{radar_beam_prediction}, and red-green-blue (RGB) images \cite{camera_beam_prediction}. Since wireless propagation characteristics are tightly coupled with the surrounding physical environment \cite{smart_factory_ba}, such sensing data can be used to predict beam index distributions and further prune the beam search space.

Early research on multi-modal sensing-aided beam management primarily targeted single-user scenarios. For instance, Feng \textit{et al.} \cite{smart_factory_ba} developed a pre-training and transfer-learning strategy to predict the optimal beam from ambient environmental information. Yang \textit{et al.} \cite{environment_semantics_beam_alignment} proposed an environment-semantics network to predict the optimal beam index and blockage state. Bian \textit{et al.} \cite{lidar_gps_beam_tracking} incorporated 3D LiDAR point clouds and location data to assist beam tracking in vehicle-to-vehicle line-of-sight (LoS) scenarios. However, LoS-oriented sensors such as cameras struggle in non-line-of-sight (NLoS) conditions, and therefore data-driven methods based only on LoS information may fail to capture dominant NLoS paths in urban environments. 
\textcolor{black}{Recent geometry-aware approaches connect propagation structure with beam management more explicitly. Kim \textit{et al.} used vision-derived user geometry to enable channel reconstruction and interference-aware beam focusing \cite{vision_aided_positioning_beam_focusing}. Wen \textit{et al.} extracted visual semantics to provide propagation-relevant scatterer geometry \cite{vision_semantics_beam_selection}, while Arnold \textit{et al.} proposed reconstructed 3D environmental models to  support beam acquisition \cite{vision_assisted_digital_twin}. Kim \textit{et al.} adopted large multi-modal models to use user positions, scatterer identities, and reflection points with sensing information and pilots to estimate channel parameters \cite{llm_beam_management}. These studies motivate a reusable  representation that preserves effective propagation structure with compact online access.}

Although multi-modal sensing has been widely studied for single-user beam management, its extension to multi-user scenarios remains underexplored. Directly transferring single-user sensing-aided methods to multi-user contexts often leads to severe performance degradation because of inter-user interference. Recently, a few works have attempted to bridge this gap. For example, Ahn \textit{et al.} \cite{vision_aided_mu_beam_selection} employed two serial deep neural networks for joint user and beam selection using captured image data, but only in LoS scenarios. Kartik \textit{et al.} \cite{multimodal_mu_beamforming} proposed a deep learning (DL) based channel estimation framework that reconstructed the complete beamspace channel using sensing data, assuming unlimited phase-shifter resolution. Nevertheless, such a framework also incurred high computational overhead because it processed raw multi-modal data. Mohammad \textit{et al.} \cite{isac_beam_management} used contextual integrated sensing and communication (ISAC) data together with a multi-modal transformer encoder for beam selection based on deep reinforcement learning (DRL). However, inter-user interference is not explicitly considered because the evaluated user distribution is sparse. Moreover, existing sensing-assisted methods often entangle multi-modal feature extraction and multi-user coordination within a unified black-box network. Separating the environmental representation from multi-user decisions makes the propagation assumptions explicit and allows the same geometric database to support different scheduled user groups.

\subsection{Main Contributions}
\textcolor{black}{To overcome the limitations of existing methods for multi-user MIMO (MU-MIMO) beam management, we develop an explicit geometric representation from
multi-modal environmental observations. Specifically, we extract site-specific propagation structures from
multi-modal observations and organize them into
interpretable geometric priors.
By exploiting the persistent geometry of dominant
reflected paths, we develop a framework that maps
raw multi-modal data into a structured database of
representative VBS locations and their associated
quantized virtual line-of-sight (VLoS) coverage regions.
Derived from virtual mirror images of the physical BS
across major reflective facades, this representation enables
analytical reconstruction of user-dependent coarse
beamspace priors for interference-aware MU-MIMO
beam management. 
While virtual-anchor methods employ mirror-image
geometry for localization and mapping
\cite{hybrid_slam}, and the channel knowledge map (CKM)-based beam-alignment
method in \cite{ckm_beam_alignment} uses
location-indexed channel parameters, the proposed
representation jointly captures the geometric origins
and spatial availability of dominant reflected paths. Specifically, the main contributions of this study
are summarized as follows:}
\begin{itemize}
\item We introduce a VBS construction scheme that converts environmental data into a sparse and interpretable radio propagation representation. Specifically, the proposed scheme extracts dominant reflective surfaces from 3D LiDAR point clouds,
constructs mirror-image VBS candidates using the
BS location, and groups them into representative
VBSs without preliminary wireless measurements. By integrating VBS locations with LoS and virtual line-of-sight (VLoS) coverage regions, the resulting database provides a compact, site-specific description of dominant propagation paths with low storage overhead.
\item Since the VBS database provides only a coarse beamspace prior, online calibration is required to account for residual geometric uncertainty and propagation variations. To this end, we propose the VBS-assisted orthogonal-pilot (VOP)-based partial beam training scheme, supported by VBS-assisted coarse channel reconstruction. The scheme infers beam-relevant parameters from geometric relationships, thereby bridging environmental geometry and online beam training while reducing the beam training overhead.
\item After partial beam training, the remaining challenge is to coordinate beam selection among multiple users under inter-user interference. We therefore propose a dual-agent dueling double deep Q-network for coordinated beam selection (DD3QN-CBS), which exploits complementary beamspace features from both the VBS-assisted coarse representation and the refined online measurements. Under a hierarchical leader-follower architecture, the BS-side and user-side agents sequentially determine their beam selection actions. Moreover, an online action masking mechanism is embedded in the BS-side agent to structurally suppress inter-user interference, accelerate training convergence, and improve beam selection performance.
\end{itemize}

The remainder of the paper is organized as follows\footnote{\textcolor{black}{Our conference study \cite{vbs_beam_alignment} introduced VBS-assisted single-user beam alignment. This work further develops the VBS representation for interference-aware MU-MIMO beam management. It integrates the proposed VOP-based partial beam training, hierarchical DD3QN-CBS coordinated selection, and evaluations of candidate pruning and location-only learning under the MU-MIMO setting.}}. Section \ref{channel_model} presents the downlink MU-MIMO beam selection problem formulation and motivates the VBS abstraction from mmWave propagation characteristics. Section \ref{vbs_construction_database} describes the offline VBS database construction from multi-modal data. Section \ref{vbs_guided_beam_management} develops the VBS-guided beam management framework, comprising coarse channel reconstruction, a VOP-based partial beam training scheme, and a DD3QN-CBS framework and implementation. Section \ref{simulation_results} presents evaluations on realistic urban layouts and comparisons with baseline strategies. Section \ref{conclusion} concludes the work.

\textcolor{black}{In this paper, $x$, $\mathbf{x}$, $\mathbf{X}$, $\mathfrak{X}$, and $\mathcal{X}$ denote a scalar, a vector, a matrix, a higher-order tensor, and a set or space, respectively. $\mathbb{R}$ and $\mathbb{C}$ denote the sets of real and complex numbers, respectively. For a matrix $\mathbf{X}$, $\mathbf{X}^{\mathsf{T}}$, $\mathbf{X}^{\mathsf{H}}$, $\mathbf{X}^{*}$, $\mathrm{Tr}(\mathbf{X})$, and $\|\mathbf{X}\|_{\mathrm{F}}$ denote its transpose, conjugate transpose, complex conjugate, trace, and Frobenius norm, respectively. The Euclidean norm of a vector $\mathbf{x}$ is denoted by $\|\mathbf{x}\|_2$, the cardinality of a set $\mathcal{X}$ by $|\mathcal{X}|$, and statistical expectation by $\mathbb{E}[\cdot]$. The symbols $\mathbf{I}_N$ and $\mathbf{1}_N$ represent the $N\times N$ identity matrix and the $N$-dimensional all-ones vector, respectively. The notation $[\mathbf{x}]_i$ denotes the $i$-th entry of a vector $\mathbf{x}$, while $[\mathbf{X}]_{i,j}$ and $[\mathbf{X}]_{\mathcal{I},\mathcal{J}}$ denote the $(i,j)$-th entry and the submatrix with rows and columns indexed by $\mathcal{I}$ and $\mathcal{J}$, respectively. For an ordered index list $\mathcal{I}$, $[\mathcal{I}]_i$ denotes its $i$-th element. The symbol $:$ denotes all indices along the corresponding dimension of a matrix or tensor.}

\section{System Model and Framework Motivation}
\label{channel_model}
This section establishes the system-level foundation for the proposed framework. We first formulate the downlink MU-MIMO beam selection problem. Based on this general formulation, we then motivate the VBS abstraction as a geometry-driven representation that connects high-frequency propagation characteristics with the subsequent VBS-guided beam management pipeline.
\begin{figure}[t!]
    \centering
    \includegraphics[width=0.999\linewidth]{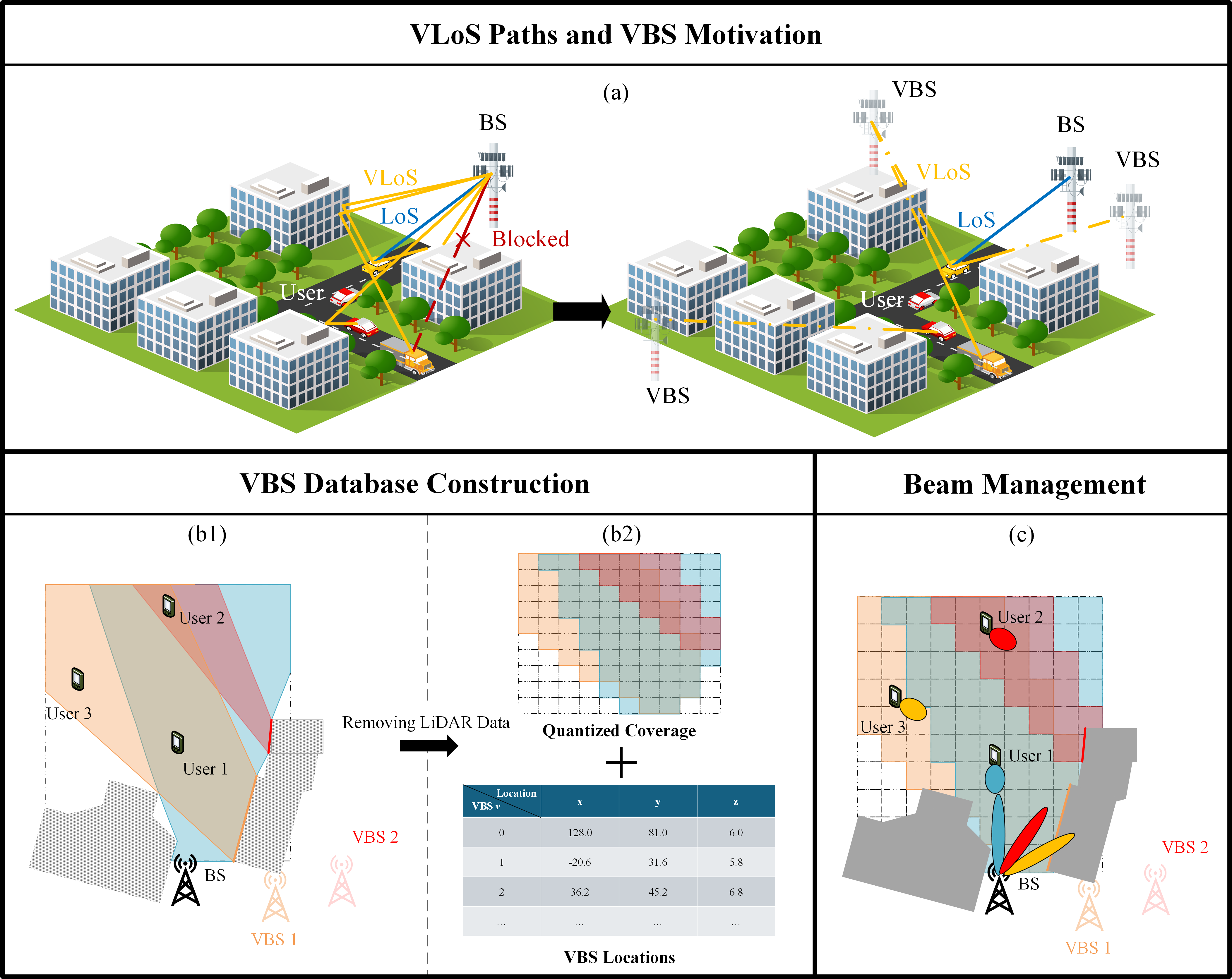}
    \caption{VBS motivation and application. (a) Building facades can generate dominant VLoS paths through single-bounce reflections when the LoS path is blocked. (b1) The VLoS coverage area can extend the effective communication area to alleviate the LoS outage problem. (b2) Constructed VBS locations and quantized coverage regions form a compact environmental abstraction where the original 3D LiDAR point clouds can be removed. (c) The resulting VBS information can support environment-aware MU-MIMO beam selection by providing spatially distinct VLoS links.}
    \label{fig:vbs_definition}
\end{figure}

\subsection{Beam Selection Problem Formulation}
\label{mu_mimo_beam_selection}
We consider a downlink MU-MIMO system where the BS is equipped with a hybrid beamforming architecture consisting of $N_\mathrm{BS}$ transmit uniform linear array (ULA) antennas and $N_\mathrm{RF}$ RF chains, serving $K$ co-scheduled users. Each user is equipped with $N_{\mathrm{UE}}$ receive ULA antennas and an analog combining architecture. The BS applies an analog precoder $\mathbf{F}_\mathrm{RF}=[\mathbf{f}_\mathrm{RF}^1, \mathbf{f}_\mathrm{RF}^2,\dotsm,\mathbf{f}_\mathrm{RF}^{N_\mathrm{RF}}]\in\mathbb{C}^{N_\mathrm{BS}\times N_\mathrm{RF}}$ and a digital baseband precoder $\mathbf{F}_\mathrm{BB}=[\mathbf{f}_\mathrm{BB}^1, \mathbf{f}_\mathrm{BB}^2,\dotsm,\mathbf{f}_\mathrm{BB}^{K}]\in\mathbb{C}^{N_\mathrm{RF}\times K}$. The $k$-th user employs an analog combining vector $\mathbf{w}_k\in\mathbb{C}^{N_{\mathrm{UE}}\times1}$, and the wireless channel from the BS to this user is denoted by $\mathbf{H}_k\in\mathbb{C}^{N_\mathrm{UE}\times N_\mathrm{BS}}$. 

Let $\mathbf{s}=[s_1,\dots,s_K]^\mathsf{T}$ be the transmitted symbol vector with $\mathbb{E}[\mathbf{s}\mathbf{s}^\mathsf{H}]=\mathbf{I}_K$. The post-combining received signal at the $k$-th user is expressed as
\begin{gather}
    y_k = \mathbf{w}_k^{\mathsf{H}}\mathbf{H}_k\mathbf{F}_\mathrm{RF}\mathbf{f}_\mathrm{BB}^k s_k + \sum_{j\neq k} \mathbf{w}_k^{\mathsf{H}}\mathbf{H}_k\mathbf{F}_\mathrm{RF}\mathbf{f}_\mathrm{BB}^j s_j + n_k,
\end{gather}
where $n_k\sim\mathcal{CN}(0,N_0W)$ represents the additive white Gaussian noise, with $N_0$ and $W$ denoting the noise power spectral density and the system bandwidth, respectively. Accordingly, the downlink signal-to-interference-plus-noise ratio (SINR) is given by
\begin{gather}
\begin{aligned}
    \mathrm{SINR}_k
    &=
    \frac{
    |\mathbf{w}_k^{\mathsf{H}}\mathbf{H}_k\mathbf{F}_\mathrm{RF}\mathbf{f}_\mathrm{BB}^k|^2
    }{
    \sum_{j\neq k}|\mathbf{w}_k^{\mathsf{H}}\mathbf{H}_k\mathbf{F}_\mathrm{RF}\mathbf{f}_\mathrm{BB}^j|^2 + N_0 W
    }.
\end{aligned}
\end{gather}
To incorporate quality of service (QoS) constraints, the system effective spectral efficiency (ESE) is evaluated by penalizing users failing an SINR threshold $\mathrm{SINR}_\mathrm{thres}$ via an indicator function $\mathbb{I}(\cdot)$, which is defined as 
\begin{gather}
    \mathrm{ESE} = \sum_{k=1}^{K} \log_2\big(1+\mathrm{SINR}_k\big) \cdot \mathbb{I}(\mathrm{SINR}_k \geq \mathrm{SINR}_\mathrm{thres}).
\end{gather}
The joint analog and digital beam selection problem maximizing the system ESE is formulated as
\begin{subequations}
\label{mu_mimo_beam_selection_problem}
\begin{alignat}{2}
    & \makebox[2.5em][l]{$\displaystyle \max_{\mathclap{\substack{\mathbf{F}_{\mathrm{RF}}, \, \{\mathbf{w}_k\}_{k=1}^{K}, \mathbf{F}_{\mathrm{BB}}}}}$} \quad && \mathrm{ESE} \label{mu_mimo_obj} \\
    & \makebox[2.5em][l]{s.t.} && \mathbf{f}_{\mathrm{RF}}^{i} \in \mathcal{F}_{\mathrm{BS}}, \quad i = 1, \dots, N_{\mathrm{RF}}, \label{mu_mimo_cons1}\\
    & && \mathbf{w}_k \in \mathcal{W}_{\mathrm{UE}}, \quad k = 1, \dots, K, \label{mu_mimo_cons2}\\
    & && \mathrm{Tr}\big( \mathbf{F}_{\mathrm{RF}} \mathbf{F}_{\mathrm{BB}} \mathbf{F}_{\mathrm{BB}}^\mathsf{H} \mathbf{F}_{\mathrm{RF}}^\mathsf{H} \big) \leq P_{\mathrm{T}}, \label{mu_mimo_cons3}
\end{alignat}
\end{subequations}
where $\mathcal{F}_\mathrm{BS}$ and $\mathcal{W}_\mathrm{UE}$ denote the discrete Fourier transform (DFT) analog beam codebooks at the BS and users, respectively, and $P_\mathrm{T}$ represents the total transmit power budget. Assuming a standard half-wavelength antenna spacing, the array response vector $\mathbf{a}(\theta; N) \in \mathbb{C}^{N \times 1}$ for an $N$-element ULA is defined as
\begin{gather}
    \mathbf{a}(\theta; N) \triangleq \frac{1}{\sqrt{N}} \left[ 1, e^{j \pi \theta}, \dots, e^{j \pi (N-1) \theta} \right]^\mathsf{T}. \label{steering_vector_def}
\end{gather}
Based on \eqref{steering_vector_def}, the analog beamforming codebooks $\mathcal{F}_\mathrm{BS}$ and $\mathcal{W}_\mathrm{UE}$ are constructed by uniformly quantizing the spatial frequency domain as
\begin{align}
    \mathcal{F}_\mathrm{BS} &\triangleq \left\{ \mathbf{a}\left( \frac{2m - 1 - N_\mathrm{BS}}{N_\mathrm{BS}}; N_\mathrm{BS} \right) \Biggm| m = 1, \dots, N_\mathrm{BS} \right\}, \\
    \mathcal{W}_\mathrm{UE} &\triangleq \left\{ \mathbf{a}\left( \frac{2n - 1 - N_\mathrm{UE}}{N_\mathrm{UE}}; N_\mathrm{UE} \right) \Biggm| n = 1, \dots, N_\mathrm{UE} \right\}.
\end{align}

In problem \eqref{mu_mimo_beam_selection_problem}, the physical implications of the mathematical objectives and hardware limitations are detailed as follows. The objective \eqref{mu_mimo_obj} targets the maximization of the system ESE. Constraints \eqref{mu_mimo_cons1} and \eqref{mu_mimo_cons2} mathematically confine the continuous search space to the discrete grids mandated by the analog phase shifter hardware resolutions. Finally, constraint \eqref{mu_mimo_cons3} accounts for the total power budget limitations. To decouple this joint combinatorial optimization tractably, the analog beamforming matrices $\mathbf{F}_\mathrm{RF}$ and $\{\mathbf{w}_k\}_{k=1}^{K}$ are first dynamically selected from the finite codebook spaces, after which the digital precoder $\mathbf{F}_\mathrm{BB}$ can be optimized conditioned on the low-dimensional effective channel matrix.

The primary challenge in solving \eqref{mu_mimo_beam_selection_problem} stems from the excessive training overhead required to acquire full CSI, coupled with the combinatorial complexity of optimizing MU-MIMO joint beam pairs. This bottleneck motivates the use of stable environmental priors that can reduce the dependence on exhaustive wireless measurements before online beam selection.

\subsection{VBS Concept and Database Representation}

\textcolor{black}{As shown in Fig.~\ref{fig:vbs_definition}(a),
wireless communication systems operating at high
frequencies, such as mmWave, face significant path
attenuation and severe blockage loss, which
frequently prevent direct LoS paths from providing
reliable coverage. Under these conditions, large
and relatively static building facades can naturally serve as
reflectors that provide auxiliary propagation
paths. This geometric property motivates the
construction of VBSs to represent reflected links
in an interpretable form.}

\textcolor{black}{
Formally, for a locally planar reflecting facade,
a raw VBS candidate is defined as the mirror image
of the physical BS with respect to the supporting
reflector plane. This virtual source represents
the geometry of a potential single-bounce specular
reflection path between the BS and a user.
For a given user location, the reflected path is
geometrically valid when the reflection point lies
on the finite facade and both physical propagation
segments are unobstructed\footnote{\textcolor{black}{
Single-bounce specular reflections constitute an
important propagation mechanism in mmWave environments.
Higher-order reflections generally incur additional
path and reflection losses, motivating the first-order
geometric approximation adopted in this work \cite{analytical_model_mmWave_nlos,mpc_characterization_mmWave}.}}.}

\textcolor{black}{
This single-bounce VBS abstraction provides a
compact geometric description of facade-assisted
propagation in high-frequency urban environments.
Ground-reflection-assisted communication has also
demonstrated the usefulness of reflected paths for
blockage mitigation \cite{TERRA}.
In this work, we also focus on facade-induced links,
whose spatial availability depends on the finite
reflector geometry and surrounding blockage.
Consequently, the virtual-source location alone
does not fully describe where the corresponding
reflected path can be used.}

\textcolor{black}{
To make this geometric abstraction directly usable
for beam management, we define a VBS database,
as shown in Fig.~\ref{fig:vbs_definition}(b1)
and Fig.~\ref{fig:vbs_definition}(b2),
that stores representative VBS locations and their
quantized coverage regions. Specifically, raw VBS
candidates with consistent geometry are grouped
into representative VBSs, as detailed in Section III.
The $v$-th representative VBS is characterized by
its location $\mathbf{o}_{\mathrm{VBS}}^{v}$
and its VLoS coverage region $\mathcal{C}_{v}$
for $v\geq 1$. For notational consistency,
the physical BS is represented by its location
$\mathbf{o}_{\mathrm{BS}}$ and LoS coverage region
$\mathcal{C}_{0}$. Together, these locations and
coverage regions characterize the geometric
availability of direct and reflected paths and
provide the information needed to construct
location-dependent beamspace priors.
The resulting database can be reused across user
locations without online processing of the original
3D LiDAR point clouds. As depicted in
Fig.~\ref{fig:vbs_definition}(c), the available
VLoS paths can provide spatially distinct beam
directions for coordinated MU-MIMO beam selection
under inter-user interference. In the following
section, we present the methodology for
constructing the VBS database.}

\section{VBS Database Construction from Multi-Modal Data}
\label{vbs_construction_database}
\begin{figure*}[t]
    \centering
    \includegraphics[width=0.99\linewidth]{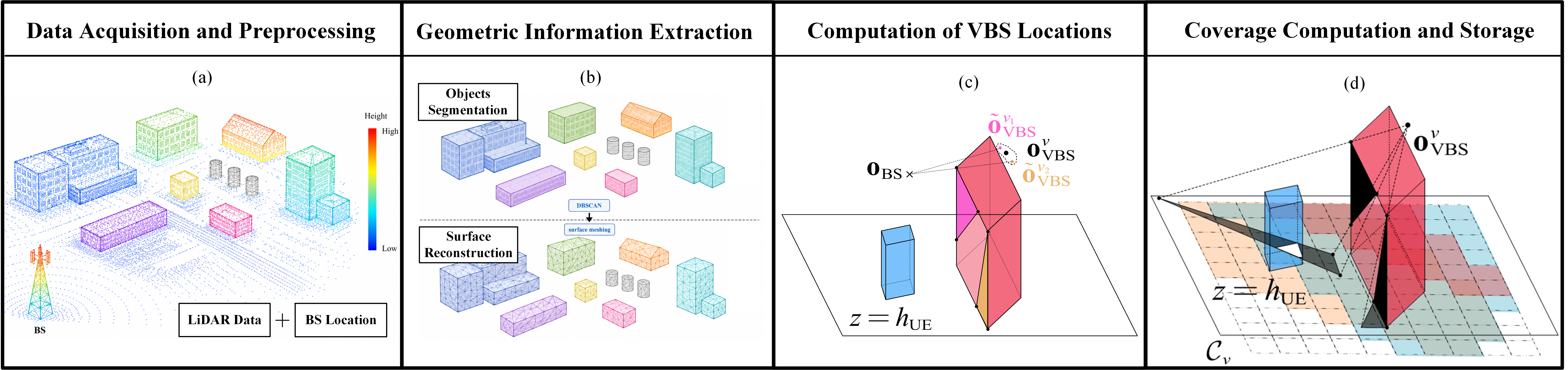}
    \caption{Offline VBS database construction. (a) Regional 3D LiDAR point cloud and BS location. (b) Building segmentation and finite surface reconstruction. (c) Mirror-source construction and clustering with centroid representatives. (d) Physical visibility validation and quantized VLoS coverage storage.}
    \label{fig:vbs_construction}
\end{figure*}

This section develops the offline construction procedure for the compact VBS database without wireless measurements. As depicted in Fig. \ref{fig:vbs_construction}, the process comprises four main stages: 3D LiDAR point cloud acquisition, geometric information extraction, VBS location computation, and coverage computation and storage. 

\subsection{Geometric Information Extraction}

\textcolor{black}{To identify major reflecting facades, the static point cloud
is first segmented into individual building objects using
the density-based spatial clustering of applications with
noise (DBSCAN) algorithm \cite{DBSCAN}.
This object-level segmentation provides geometrically
coherent regions for subsequent surface reconstruction.
For the $m$-th segmented object, surface reconstruction
\cite{alpha_shape} and simplification \cite{QEM}
produce $N_m$ triangular meshes.
Each triangular mesh $\mathcal{T}_j$ is defined by its
three vertices $\mathbf{v}_j^k\in\mathbb{R}^{3\times1}$
($k=1,2,3$).
Its centroid is given by
$\mathbf{v}_j^{\mathrm{c}}
=\frac{1}{3}\sum_{k=1}^{3}\mathbf{v}_j^k$,
while its surface normal $\mathbf{n}_j$ is computed
from the cross product of two edge vectors.
These geometric quantities characterize the local
surface positions and orientations.}

\textcolor{black}{Building facades may exhibit local curvature and
surface reconstruction variations.
Neighboring triangular meshes with consistent
geometric orientations are therefore organized
into finite, locally planar facade segments
$\mathcal{L}_i$.
Each segment is associated with a representative
reflector plane characterized by a reference point
$\mathbf{v}_i^{\mathrm{c}}$ and normal vector
$\mathbf{g}_i$, while its constituent triangles
retain the finite surface geometry for subsequent
intersection and visibility calculations.
The extracted facade segments provide the
geometric primitives for generating raw VBS
candidates through the mirror principle.
Their physical validity is examined in the
following subsection.}

\subsection{Determination of VBS Locations}
\label{vbs_location_computation}

\textcolor{black}{Based on the finite facade segments extracted in the previous subsection, candidate VBS locations are constructed by mirroring the physical BS across their corresponding reflector planes. Before generating raw VBS candidates, we examine the reflector orientation and BS-side visibility to identify feasible reflecting facades. The user-dependent reflected paths are subsequently validated during VLoS coverage computation.}

\textcolor{black}{The orientation condition identifies building facades facing the BS. For the $i$-th facade segment, let $\tilde{\mathbf g}_i=\mathbf g_i/\|\mathbf g_i\|_2$ denote its unit outward normal, and let $\tilde{\mathbf p}_{i,\mathrm{BS}}=(\mathbf o_{\mathrm{BS}}-\mathbf v_i^\mathrm{c})/\|\mathbf o_{\mathrm{BS}}-\mathbf v_i^\mathrm{c}\|_2$ denote the unit direction from the facade center to the BS. The orientation condition is expressed as
\begin{equation}
\tilde{\mathbf g}_i^{\mathsf T}
\tilde{\mathbf p}_{i,\mathrm{BS}}>\tau_{\mathrm{ori}},
\end{equation}
where $\tau_{\mathrm{ori}}=0.02$ excludes near-grazing incidence. The outward normal directions are determined from the consistently oriented building boundaries.}

\textcolor{black}{The unobstructed path condition determines whether
the finite facade is visible from the BS.
For each facade segment $\mathcal{L}_i$, the three vertices
of each constituent triangular element serve as representative
points $\mathbf{q}$.
The intersection indicator
$\mathrm{Intersect}
(\mathbf{o}_{\mathrm{BS}},\mathbf{q},\mathcal{T}_j)$
equals one when the segment from the BS to
$\mathbf{q}$ intersects mesh $\mathcal{T}_j$,
and zero otherwise.
It is evaluated using the M\"oller-Trumbore
ray-triangle intersection algorithm
\cite{ray_tracing}.
A facade satisfies the visibility condition if at least one representative point $\mathbf{q}$ on $\mathcal{L}_i$ has an unobstructed path from the BS, namely,}
{\color{black}
\begin{equation}
    \sum_{\mathcal{T}_j\nsubseteq\mathcal{L}_i}
    \mathrm{Intersect}
    (\mathbf{o}_{\mathrm{BS}},\mathbf{q},\mathcal{T}_j)
    =0.
\end{equation}
}
\textcolor{black}{This condition identifies facades with at least
one unobstructed incident path from the BS.
The user-dependent reflection geometry is
further validated during VLoS coverage computation.}

\textcolor{black}{Upon satisfying these conditions, for each valid facade segment, the raw VBS location is obtained
by mirroring the physical BS across its supporting plane,}
{\color{black}
\begin{equation}
    \tilde{\mathbf{o}}_{\mathrm{VBS}}^{i}
    =
    \mathbf{o}_{\mathrm{BS}}
    -2d_i\tilde{\mathbf{g}}_i,
\end{equation}}
\textcolor{black}{where
$d_i=\tilde{\mathbf{g}}_i^\mathsf{T}
(\mathbf{o}_{\mathrm{BS}}-\mathbf{v}_i^{\mathrm{c}})$
is the signed distance from the BS to the reflector plane.
Reversing the normal direction also reverses the sign of
$d_i$, leaving the resulting VBS location unchanged.}

\textcolor{black}{In practice, LiDAR noise and surface reconstruction errors
may cause different segments of the same physical facade
to produce slightly displaced raw VBS locations.
We therefore apply hierarchical density-based spatial
clustering of applications with noise (HDBSCAN)
\cite{HDBSCAN} to group geometrically consistent raw VBS
candidates into clusters.
Reflector-orientation consistency is also considered
to avoid grouping candidates associated with
geometrically different facades. 
For the $v$-th cluster, the representative VBS location
$\mathbf{o}_{\mathrm{VBS}}^v$ is computed as the centroid
of its constituent raw VBS locations.
The associated facade geometries are retained for
subsequent reflection-point and visibility calculations,
as illustrated in Fig.~\ref{fig:vbs_construction}(c). In addition to building facades, the ground is
approximated as a single horizontal reflecting plane,
yielding one additional VBS through the same mirror-image
construction. }

\subsection{Coverage Computation and Database Storage}
\label{coverage_computation_storage}

\textcolor{black}{Following the construction and clustering of valid raw VBS candidates, we determine the coverage regions associated with the physical BS and the representative VBSs.}
\textcolor{black}{To efficiently store this spatial coverage information, the service region is discretized into a $D_{\mathrm{x}} \times D_{\mathrm{y}}$ grid at the user height slice $z = h_{\mathrm{UE}}$, as illustrated in Fig.~\ref{fig:vbs_construction}(d).}
The spatial coverage associated with the physical BS and each clustered VBS is characterized by a discrete coordinate set $\mathcal{C}_v$.
Specifically, $\mathcal{C}_0$ aggregates the grid points providing direct LoS service, while each $\mathcal{C}_v$ ($v \geq 1$) \textcolor{black}{contains the grid points supporting valid reflected links associated with the $v$-th VBS.}

\textcolor{black}{To populate these coverage sets, each grid point $\mathbf{o}_{\mathrm{g}}$ is evaluated against the reconstructed environment.
For the LoS coverage $\mathcal{C}_0$, a grid point is included when the direct path from $\mathbf{o}_{\mathrm{BS}}$ to $\mathbf{o}_{\mathrm{g}}$ is unobstructed.
For the VLoS coverage $\mathcal{C}_v$ ($v\geq1$), the finite reflector geometries associated with the constituent raw VBS candidates are retained after clustering.
For each constituent reflector, the image-source geometry is used to determine a candidate reflection point on its supporting plane.
A valid reflected path is identified when the reflection point lies within the modeled reflecting surface and both physical propagation segments, from the BS to the reflection point and from the reflection point to the grid point, are unobstructed.
The grid point is included in $\mathcal{C}_v$ if at least one constituent reflector provides such a path, as illustrated in Fig.~\ref{fig:vbs_construction}(d).}

\textcolor{black}{The completed VBS database stores the physical BS entry $\{\mathbf{o}_{\mathrm{BS}},\mathcal{C}_0\}$ and the representative VBS entries $\{\mathbf{o}_{\mathrm{VBS}}^{v},\mathcal{C}_{v}\}_{v\geq1}$.
The quantized coverage sets characterize spatial path availability, while the stored BS and VBS locations enable analytical reconstruction of user-dependent propagation geometry and beamspace priors.
This representation separates offline environmental geometry processing from online beam management, which operates on the compact database without repeatedly processing the original environmental observations.
When the BS location changes within an unchanged static environment, the acquired environmental geometry can be reused to reconstruct the new corresponding VBS locations and coverage regions.}

\section{VBS-guided MU-MIMO Beam Management}
\label{vbs_guided_beam_management}

In this section, we present the VBS-guided MU-MIMO beam management framework by developing the VBS-assisted coarse channel reconstruction, VOP-based partial beam training, and DD3QN-CBS algorithm. At the framework level, the offline VBS database described in Section \ref{vbs_construction_database} is converted into a VBS-assisted coarse beamspace representation, which is subsequently calibrated through online partial beam training for interference-aware multi-user coordination.

As illustrated in Fig. \ref{fig:decision_auxiliary_beam}, the proposed framework consists of three sequential stages. First, the spatial relations among the BS, VBSs, and users are mapped into the channel domain, producing a coarse beamspace representation. Second, since this representation is derived from the VBS database and may be affected by LiDAR noise, reconstruction errors, and grid quantization, a VOP-based partial beam training scheme is designed to measure and refine the most promising candidate beam entries with limited online overhead. Finally, the refined online measurements and the VBS-assisted coarse beamspace representation are jointly incorporated into the DD3QN-CBS algorithm to perform coordinated MU-MIMO beam selection under multi-user interference. The details of these three stages are presented as follows.

\subsection{VBS-assisted Coarse Channel Reconstruction}

This subsection details the transformation of the offline VBS database into the coarse channel estimates. By extracting path parameters, such as angle-of-arrival (AoA) and angle-of-departure (AoD), from the stored locations and coverage regions, the system constructs this representation to guide subsequent beam training and selection. 
\subsubsection{Possible AoA and AoD Computation}
\label{aoa_aod_estimation}
Given the user location and the VBS database, the BS first identifies the available LoS or VLoS paths through $\mathcal{C}_{0}$ and $\{\mathcal{C}_{v}\}_{v\geq 1}$. \textcolor{black}{The representative VBS geometry estimates a reflection point for each active VLoS path and maps the available paths into coarse angular coordinates.} Specifically, for a clustered VBS $v \geq 1$, let $\mathbf{p}_{\mathrm{BS}}^{v} = \mathbf{o}_{\mathrm{ref}}^{v} - \mathbf{o}_{\mathrm{BS}}$ and $\mathbf{p}_{\mathrm{UE}}^{v} = \mathbf{o}_{\mathrm{ref}}^{v} - \mathbf{o}_{\mathrm{UE}}$ \textcolor{black}{denote the displacement vectors from the BS and the user to the reflection point associated with VBS $v$, respectively.} The coordinates of the reflection point $\mathbf{o}_{\mathrm{ref}}^{v}$ can be computed \textcolor{black}{from the representative VBS location} $\mathbf{o}_{\mathrm{VBS}}^{v}$ via geometric mirror relations as
\begin{equation}
\mathbf{o}_{\mathrm{ref}}^{v} = \mathbf{o}_{\mathrm{VBS}}^{v} + \frac{ \lVert\mathbf{o}_{\mathrm{BS}} - \mathbf{o}_{\mathrm{VBS}}^{v}\rVert_2^2 \left( \mathbf{o}_{\mathrm{UE}} - \mathbf{o}_{\mathrm{VBS}}^{v} \right) }{ 2 \left( \mathbf{o}_{\mathrm{BS}} - \mathbf{o}_{\mathrm{VBS}}^{v} \right)^\top \left( \mathbf{o}_{\mathrm{UE}} - \mathbf{o}_{\mathrm{VBS}}^{v} \right) }.
\end{equation}
For the LoS path, when it is present, the corresponding displacement vectors are $\mathbf{p}_{\mathrm{BS}}^{0} = \mathbf{o}_{\mathrm{UE}} - \mathbf{o}_{\mathrm{BS}}$ and $\mathbf{p}_{\mathrm{UE}}^{0} = \mathbf{o}_{\mathrm{BS}} - \mathbf{o}_{\mathrm{UE}}$. We introduce a unified geometric mapping $ f[\cdot;\cdot, \cdot] $ defined as 
\begin{gather}
    f[\mathbf{p};\Delta\phi,\Delta\psi] \triangleq
    \begin{bmatrix}
        \operatorname{atan2}\!\left([\mathbf{p}]_2,[\mathbf{p}]_1\right) + \Delta\phi \\
        \arccos\!\left(\frac{[\mathbf{p}]_3}{\|\mathbf{p}\|_2}\right) + \Delta\psi \\
        \|\mathbf{p}\|_2
    \end{bmatrix},
    \label{geometric_mapping}
\end{gather}
where the array azimuth and elevation offsets at the user and BS are denoted as $\Delta\phi^{\mathrm{r}},\Delta\phi^\mathrm{t}$ and $\Delta\psi^{\mathrm{r}},\Delta\psi^\mathrm{t}$, respectively. Using \eqref{geometric_mapping}, the user-side and BS-side geometric parameters for both VLoS paths and the LoS path are obtained as $\bigl[\hat{\phi}_{v}^\mathrm{r},\,\hat{\psi}_{v}^\mathrm{r},\,\hat{r}_{v}^\mathrm{r}\bigr]^\top = f[\mathbf{p}_\mathrm{UE}^{v};\Delta\phi^\mathrm{r},\Delta\psi^\mathrm{r}]$, $\bigl[\hat{\phi}_{v}^\mathrm{t},\,\hat{\psi}_{v}^\mathrm{t},\,\hat{r}_{v}^\mathrm{t}\bigr]^\top = f[\mathbf{p}_\mathrm{BS}^{v};\Delta\phi^\mathrm{t},\Delta\psi^\mathrm{t}]$, $\bigl[\hat{\phi}_{0}^\mathrm{r},\,\hat{\psi}_{0}^\mathrm{r},\,\hat{r}_{0}^\mathrm{r}\bigr]^\top = f[\mathbf{p}_\mathrm{UE}^{0};\Delta\phi^\mathrm{r},\Delta\psi^\mathrm{r}]$, and $\bigl[\hat{\phi}_{0}^\mathrm{t},\,\hat{\psi}_{0}^\mathrm{t},\,\hat{r}_{0}^\mathrm{t}\bigr]^\top = f[\mathbf{p}_\mathrm{BS}^{0};\Delta\phi^\mathrm{t},\Delta\psi^\mathrm{t}]$, 
where $\hat{\phi}_{v}^\mathrm{r}$ and $\hat{\psi}_{v}^\mathrm{r}$ represent the estimated user-side azimuth and elevation angles for path $v$, respectively, while $\hat{\phi}_{v}^\mathrm{t}$ and $\hat{\psi}_{v}^\mathrm{t}$ represent the corresponding BS-side angles. The distances from the reflection point to the user and BS are denoted by $\hat{r}_{v}^\mathrm{r}$ and $\hat{r}_{v}^\mathrm{t}$, respectively, while $\hat{r}_{0}^\mathrm{r} = \hat{r}_{0}^\mathrm{t}$ represents the LoS distance between the user and BS.

These geometric quantities determine the plausible departure and arrival directions. With the path geometry fixed, the next step is to assign each candidate path a reasonable path gain before synthesizing an initial channel estimate.

\subsubsection{Path Loss Computation}

\textcolor{black}{The VBS database captures the geometry of reflected
propagation paths without material-specific RF
characterization. Accordingly, path attenuation is
approximated using free-space propagation loss and
a common effective reflection loss for VLoS paths.
The resulting path amplitudes provide coarse
propagation information for subsequent beamspace
reconstruction. Let
$d_0=\|\mathbf{p}_\mathrm{BS}^{0}\|_2$ and
$d_v=\|\mathbf{p}_\mathrm{UE}^{v}\|_2+
\|\mathbf{p}_\mathrm{BS}^{v}\|_2$ denote the
propagation distances for LoS and VLoS paths,
respectively. Their path losses are expressed as}
\begin{equation}
\begin{aligned}
    \mathrm{PL}_{0}
    &= 20\log_{10}\left(\frac{4\pi f_\mathrm{c}}{c}\right)
       +20\log_{10}(d_0), \\
    \mathrm{PL}_{v}
    &= 20\log_{10}\left(\frac{4\pi f_\mathrm{c}}{c}\right)
       +20\log_{10}(d_v)+\Gamma,
\end{aligned}
\label{total_path_loss}
\end{equation}
\textcolor{black}{where $f_\mathrm{c}$ is the carrier frequency,
$c$ is the speed of light, and $\Gamma$ denotes
the nominal effective reflection loss in dB.
The corresponding path-amplitude coefficient
is approximated as
$\hat{\beta}_v=10^{-\mathrm{PL}_v/20}$.}

\textcolor{black}{The common reflection loss introduces an
amplitude scaling factor
$\rho(\Gamma)=10^{-\Gamma/20}$ for reflected
paths. For a user with VLoS-only propagation,
this factor uniformly scales the reflected
beamspace component and preserves its
beam-magnitude ranking. When LoS and VLoS
paths coexist, $\Gamma$ adjusts their relative
amplitudes and may affect the resulting
beamspace ranking.}

\subsubsection{Coarse Channel Reconstruction}
Building upon the geometric information derived in the previous subsections, this stage synthesizes a coarse channel matrix to identify the dominant spatial directions. Due to point-cloud measurement noise and quantization errors in the stored coverage regions, using a single nearest grid point may omit useful propagation paths. To make the retrieval process less sensitive to such errors, we relax the search criteria by mapping the user location to two K-nearest-neighbor (KNN) grid-index sets, including $\mathcal{K}_0$ for the direct LoS path and $\mathcal{K}_v$ for the reflected VLoS paths. In practice, these two neighborhoods can have different sizes to reflect the different uncertainty levels of direct and reflected coverage. The candidate path set $\mathcal{V}_\mathrm{UE}$ is then formed as
\begin{gather}
    \mathcal{V}_\mathrm{UE}
    = \{v \mid \mathcal{C}_{v} \cap \mathcal{K}_v \neq \emptyset,\ v\geq 0\}.
\end{gather}
Given $\mathcal{V}_\mathrm{UE}$, we reconstruct the coarse channel $\hat{\mathbf{H}}$ by aggregating all identified paths through a unified summation, which can be expressed as
\begin{equation}
\begin{aligned}
    \hat{\mathbf{H}}
    &= \sqrt{N_\mathrm{BS}N_\mathrm{UE}}
    \sum_{v\in \mathcal{V}_\mathrm{UE}}\hat{\beta}_{v}e^{j\Xi_v}
    \mathbf{a}(\hat{\nu}_{v}; N_\mathrm{UE})
    \mathbf{a}(\hat{\mu}_{v}; N_\mathrm{BS})^\mathsf{H},
\end{aligned}
\label{reconstructed_channel}
\end{equation}
where the array response vector $\mathbf{a}(\cdot;\cdot)$ follows the definition in \eqref{steering_vector_def}, and the estimated BS-side and user-side spatial frequencies for path $v$ are defined as functions of the physical geometric angles, i.e., $\hat{\mu}_{v} = \cos(\hat{\phi}_{v}^\mathrm{t}) \sin(\hat{\psi}_{v}^\mathrm{t})$ and $\hat{\nu}_{v} = \cos(\hat{\phi}_{v}^\mathrm{r}) \sin(\hat{\psi}_{v}^\mathrm{r})$. For the LoS path term, the phase is deterministic and set as
$\Xi_0 = -\frac{2\pi f_\mathrm{c}}{c} \|\mathbf{p}_\mathrm{BS}^{0}\|_2$, where $f_\mathrm{c}$ is the carrier frequency and $c$ is the speed of light. 
For reflected paths, $\Xi_v\sim\mathcal{U}(0,2\pi)$ denotes independent random phases, consistent with the uncorrelated-scattering approximation used in statistical reflection modeling \cite{analytical_model_mmWave_nlos}. By preserving these core spatial signatures, $\hat{\mathbf{H}}$ provides a coarse channel representation for ranking beam pairs and narrowing the search space in the subsequent partial training phase.

\subsection{VOP-based Partial Beam Training}
\label{vop_scheme}

The VBS-assisted coarse channel reconstruction in the previous subsection provides an initial channel estimate for each user based on the available LoS and VLoS geometric paths. However, this estimate may deviate from the actual wireless channel due to LiDAR noise, grid quantization, unknown reflection properties, and unmodeled scattering components. To refine the coarse estimate without considerable beam sweeping overhead, this subsection develops a VOP-based partial beam training scheme, where the reconstructed coarse channels are projected onto the BS-side and user-side beam codebooks to identify beam pairs that are likely to contain dominant channel energy. Online pilot training is then performed only over these promising directions, so the coarse reconstruction provides the measurement plan while the pilot responses provide channel-aware calibration.

A key design issue is that online calibration and final coordinated beam selection require different levels of pruning. The measured beam set should be small to control training overhead, whereas the feasible set for final beam selection should remain broader to retain optimal joint beam configurations. Therefore, we adopt a two-level beam subset design: the search subsets specify the beam indices measured online, while the candidate subsets define the feasible beam indices for later beam selection.

\subsubsection{Beam Search and Candidate Subset Construction}
This stage constructs the two-level beam subsets by first mapping the reconstructed coarse channel estimates into the codebook beam domain. For the $k$-th user, the coarse channel estimate $\hat{\mathbf{H}}_k$ is projected onto the BS-side and user-side DFT codebooks, yielding the coarse beamspace gain matrix $ \hat{\mathbf{G}}_k = \mathbf{U}_\mathrm{UE}^\mathsf{H}\hat{\mathbf{H}}_k\mathbf{U}_\mathrm{BS} $. The BS-side and user-side DFT codebook matrices $\mathbf{U}_\mathrm{BS}$ and $\mathbf{U}_\mathrm{UE}$ are obtained by stacking the codebook vectors in $\mathcal{F}_\mathrm{BS}$ and $\mathcal{W}_\mathrm{UE}$ defined in Section \ref{mu_mimo_beam_selection}, respectively. They are defined as
\begin{align}
    \mathbf{U}_\mathrm{BS}
    &\triangleq
    \left[
    \mathbf{a}\!\left(\frac{1-N_\mathrm{BS}}{N_\mathrm{BS}};N_\mathrm{BS}\right),
    \ldots,
    \mathbf{a}\!\left(\frac{N_\mathrm{BS}-1}{N_\mathrm{BS}};N_\mathrm{BS}\right)
    \right], \\
    \mathbf{U}_\mathrm{UE}
    &\triangleq
    \left[
    \mathbf{a}\!\left(\frac{1-N_\mathrm{UE}}{N_\mathrm{UE}};N_\mathrm{UE}\right),
    \ldots,
    \mathbf{a}\!\left(\frac{N_\mathrm{UE}-1}{N_\mathrm{UE}};N_\mathrm{UE}\right)
    \right].
\end{align}
\textcolor{black}{The matrix $\hat{\mathbf{G}}_k\in\mathbb{C}^{N_{\mathrm{UE}}\times N_{\mathrm{BS}}}$ uses user beams as rows and BS beams as columns. Thus, $[\hat{\mathbf{G}}_k]_{j,i}$ denotes the predicted gain for BS beam $i$ and user beam $j$.} Therefore, subset construction is based on the magnitude ordering of the entries in $\{\hat{\mathbf{G}}_k\}_{k=1}^{K}$. The goal is to form ordered BS-side and user-side beam lists from which both the candidate subsets and the search subsets can be extracted.

Algorithm \ref{beam_training_subset_selection} implements this ordering through a greedy selection procedure. At each iteration, the triplet $(k,i,j)$ with the largest $|[\hat{\mathbf{G}}_k]_{j,i}|$ is selected, where $i$ and $j$ denote the BS-side and user-side beam indices, respectively. If the selected BS-side index has not been included and the target BS candidate size has not been reached, it is appended to the ordered BS-side list. Similarly, the selected user-side index is appended to the ordered list of user $k$ when it is new and the target user candidate size has not been reached. The selected entry is then removed from further consideration, and the process continues until the required candidate sizes are reached.

After the ordered lists are obtained, the two-level subsets are generated by taking different prefixes of the same lists. The longer prefixes form the candidate subsets $\mathcal{I}_\mathrm{BS}$ and $\mathcal{I}_\mathrm{UE}^k$, with sizes $\tilde{N}_\mathrm{BS}$ and $\tilde{N}_\mathrm{UE}$, which define the feasible beam indices for later beam selection. The shorter prefixes form the search subsets $\bar{\mathcal{I}}_\mathrm{BS}$ and $\bar{\mathcal{I}}_\mathrm{UE}^k$, with sizes $\bar{N}_\mathrm{BS}$ and $\bar{N}_\mathrm{UE}$, which are measured during VOP-based partial beam training. 

\begin{algorithm}[t]
\caption{Beam Search and Candidate Subset Construction}
\label{beam_training_subset_selection}
\begin{algorithmic}[1]
\State \textbf{Input:} Coarse beamspace $\{\hat{\mathbf{G}}_k\}_{k=1}^K$, target sizes $\tilde{N}_\mathrm{BS}, \tilde{N}_\mathrm{UE}, \bar{N}_\mathrm{BS}, \bar{N}_\mathrm{UE}$.
\State \textbf{Output:} $\mathcal{I}_\mathrm{BS}, \mathcal{I}_\mathrm{UE}^k$ and $\bar{\mathcal{I}}_\mathrm{BS}, \bar{\mathcal{I}}_\mathrm{UE}^k$.
\State Initialize ordered lists $\mathcal{I}_\mathrm{BS} = \emptyset, \mathcal{I}_\mathrm{UE}^k = \emptyset, \forall k$. 
\State Create copies $\hat{\mathbf{G}}_k^\mathrm{copy} = \hat{\mathbf{G}}_k$.
\While{$|\mathcal{I}_\mathrm{BS}| < \tilde{N}_\mathrm{BS}$ or $\exists k:|\mathcal{I}_\mathrm{UE}^k| < \tilde{N}_\mathrm{UE}$}
    \State \textcolor{black}{Find $(k_\mathrm{s},i,j)=\arg\max_{k,i,j}|[\hat{\mathbf{G}}_k^\mathrm{copy}]_{j,i}|$ over remaining entries.}
    \If {$i \notin \mathcal{I}_\mathrm{BS}$ and $|\mathcal{I}_\mathrm{BS}| < \tilde{N}_\mathrm{BS}$}
        \State Append $i$ to $\mathcal{I}_\mathrm{BS}$.
    \EndIf
    \If {$j \notin \mathcal{I}_\mathrm{UE}^{k_\mathrm{s}}$ and $|\mathcal{I}_\mathrm{UE}^{k_\mathrm{s}}| < \tilde{N}_\mathrm{UE}$}
        \State Append $j$ to $\mathcal{I}_\mathrm{UE}^{k_\mathrm{s}}$.
    \EndIf
    \State \textcolor{black}{Exclude entry $(j,i)$ of user $k_\mathrm{s}$ from subsequent maximizations.}
\EndWhile
\State Select the first $\bar{N}_\mathrm{BS}$ entries of $\mathcal{I}_\mathrm{BS}$ to form $\bar{\mathcal{I}}_\mathrm{BS}$, and select the first $\bar{N}_\mathrm{UE}$ entries of each $\mathcal{I}_\mathrm{UE}^k$ to form $\bar{\mathcal{I}}_\mathrm{UE}^k$. 
\end{algorithmic}
\end{algorithm}

\subsubsection{Partial Beam Measurements and Feature Synthesis}
After the beam search subsets have been selected, the remaining task is to measure these indexed entries across the selected beam configurations. Sequential user training over the reduced subsets requires $\lceil\bar{N}_\mathrm{BS}/N_\mathrm{RF}\rceil\bar{N}_\mathrm{UE}K$ beam sweeping rounds, where $\lceil \cdot \rceil$ denotes the ceiling function. The proposed VOP-based scheme combines VBS-guided subset construction with orthogonal-pilot (OP)-based concurrent pilot transmission. VBS-guided subset selection reduces the measured BS-side and user-side beam dimensions from $N_\mathrm{BS}$ and $N_\mathrm{UE}$ to $\bar{N}_\mathrm{BS}$ and $\bar{N}_\mathrm{UE}$. Orthogonal pilots remove the user-wise sequential factor by supporting distinguishable concurrent uplink transmission from all users. Exhaustive sequential training requires $\lceil N_\mathrm{BS}/N_\mathrm{RF}\rceil N_\mathrm{UE}K$ beam sweeping rounds \cite{tutorial_bm}, while full-codebook concurrent OP-based training requires $\lceil N_\mathrm{BS}/N_\mathrm{RF}\rceil N_\mathrm{UE}$ beam sweeping rounds \cite{interlaced_beam_training_allocation}. The proposed scheme requires $\lceil \bar{N}_\mathrm{BS}/N_\mathrm{RF}\rceil \bar{N}_\mathrm{UE}$ beam sweeping rounds. Each round uses one BS beam group and one user-side beam configuration. The pilot length within a round is specified separately by $\tau\geq K$.

For compact notation, let $\bar{\mathbf{b}}=[\bar{b}_1,\ldots,\bar{b}_{\bar{N}_\mathrm{BS}}]$ denote the ordered BS-side indices in $\bar{\mathcal{I}}_\mathrm{BS}$, and let $\bar{\mathbf{u}}^k=[\bar{u}_1^k,\ldots,\bar{u}_{\bar{N}_\mathrm{UE}}^k]$ denote the ordered user-side indices in $\bar{\mathcal{I}}_\mathrm{UE}^k$. The $m$-th BS training group is defined as
\begin{gather}
    \mathcal{B}_m
    =
    \{\bar{b}_\ell \mid \ell=(m-1)N_\mathrm{RF}+1,\ldots,mN_\mathrm{RF}\},
\end{gather}
with the last group truncated if necessary. 
Specifically, we adopt an uplink pilot training procedure.
In each training slot $(m,n)$, all $K$ users concurrently
transmit their pilot symbols using their respective
\textcolor{black}{uplink transmit beam vectors
\begin{gather}
    \mathbf{t}_k^{(n)} = [\mathbf{U}_\mathrm{UE}]_{:,\bar{u}_n^k}^{*}.
\end{gather}
Here, $(\cdot)^*$ denotes element-wise
complex conjugation. The BS receives the overlapped signals using the analog combiner
\begin{gather}
    \mathbf{W}_\mathrm{RF}^{(m)}
    =
    [\mathbf{U}_\mathrm{BS}]_{:,\mathcal{B}_m}^{*}.
\end{gather}}
Orthogonal pilots $\bm{\upphi}_k \in \mathbb{C}^{1 \times \tau}$
are employed to differentiate the uplink pilot signals
from different users at the BS, satisfying
$\bm{\upphi}_k \bm{\upphi}_j^\mathsf{H} = 1$
if $k=j$ and $0$ otherwise, where $\tau \geq K$
is the pilot length.
\textcolor{black}{The VOP-based training operates in a time-division duplex (TDD) system with calibrated RF front ends, where uplink pilot transmission and subsequent downlink communication occur within the channel coherence interval. Under the adopted signal convention, the reciprocal uplink channel is $\mathbf{H}_k^{\mathsf{T}}$, and the conjugated DFT codewords preserve the corresponding downlink beam directions. Accordingly, the received signal matrix $\mathbf{Y}^{(m,n)}\in\mathbb{C}^{N_{\mathrm{RF}}\times\tau}$ is expressed as}
{\color{black}
\begin{equation}
    \mathbf{Y}^{(m,n)}
    =
    \sum_{k=1}^{K}
    \sqrt{\tau P_\mathrm{p}}
    (\mathbf{W}_\mathrm{RF}^{(m)})^\mathsf{H}
    \mathbf{H}_k^\mathsf{T}
    \mathbf{t}_k^{(n)}
    \bm{\upphi}_k
    +
    (\mathbf{W}_\mathrm{RF}^{(m)})^\mathsf{H}
    \mathbf{N}.
\end{equation}}
where $P_\mathrm{p}$ is the pilot transmit power
and $\mathbf{N}$ is the additive white Gaussian noise
with entries following $\mathcal{CN}(0,N_0W)$.
By correlating the received signal with the corresponding
pilot $\bm{\upphi}_k$, the BS leverages this orthogonality
to suppress inter-user interference and decode the
measurement vector
$\mathbf{g}_k^{(m,n)}
\in \mathbb{C}^{N_\mathrm{RF} \times 1}$
for the $k$-th user, which can be expressed as
{\color{black}
\begin{gather}
\begin{aligned}
    \mathbf{g}_k^{(m,n)}
    &=
    \frac{1}{\sqrt{\tau P_\mathrm{p}}}
    \mathbf{Y}^{(m,n)}
    \bm{\upphi}_k^\mathsf{H}
    \\
    &=
    (\mathbf{W}_\mathrm{RF}^{(m)})^\mathsf{H}
    \mathbf{H}_k^\mathsf{T}
    \mathbf{t}_k^{(n)}
    +
    \frac{1}{\sqrt{\tau P_\mathrm{p}}}
    (\mathbf{W}_\mathrm{RF}^{(m)})^\mathsf{H}
    \mathbf{N}
    \bm{\upphi}_k^\mathsf{H}.
\end{aligned}
\end{gather}}

After completing the measurements across all training slots, the decoded vectors $\mathbf{g}_k^{(m,n)}$ are assigned to populate a measured beamspace matrix $\tilde{\mathbf{G}}_k \in \mathbb{C}^{N_\mathrm{UE} \times N_\mathrm{BS}}$, which is initialized to zero. The entries are filled according to the measured indices, such that the $(j,i)$-th entry of $\tilde{\mathbf{G}}_k$ corresponding to the measured BS and user beams is assigned the value of $\mathbf{g}_k^{(m,n)}$. Formally, this mapping can be expressed as
\begin{equation}
    [\tilde{\mathbf{G}}_k]_{\bar{u}_n^k, \mathcal{B}_m} = (\mathbf{g}_k^{(m,n)})^\mathsf{T}.
\end{equation}
Finally, we extract the effective beamspace submatrices corresponding to the candidate subsets $\mathcal{I}_\mathrm{BS}$ and $\mathcal{I}_\mathrm{UE}^k$ to obtain a compact beamspace feature representation. We denote the online beamspace feature and the coarse beamspace feature as
\begin{gather}
    \mathbf{X}_{k}^{\mathrm{meas}}
    =
    [\tilde{\mathbf{G}}_k]_{\mathcal{I}_\mathrm{UE}^k,\mathcal{I}_\mathrm{BS}},
    \quad
    \mathbf{X}_{k}^{\mathrm{coar}}
    =
    [\hat{\mathbf{G}}_k]_{\mathcal{I}_\mathrm{UE}^k,\mathcal{I}_\mathrm{BS}}.
\end{gather}
The coarse and measured beamspace features provide complementary propagation information for subsequent coordinated beam selection.

\subsection{VBS-guided DD3QN-CBS}
\label{rl_beam_selection}

This subsection explains how the coarse beamspace features and refined online measurements are used to make the final beam decisions through the proposed VBS-guided DD3QN-CBS algorithm. After candidate pruning, the task is reduced from full-codebook search to selecting a joint beam configuration from the candidate spaces. This reduced problem is still nontrivial because the beam choices of different users are coupled through inter-user interference, and the ESE objective in \eqref{mu_mimo_obj} is available only after the BS beams, user beams, and digital precoder are jointly determined.

Direct exhaustive search is still expensive since the BS must select $N_\mathrm{RF}$ beams from $\tilde{N}_\mathrm{BS}$ candidates and the $K$ users jointly contribute ${\tilde{N}_\mathrm{UE}}^K$ user-side choices, leading to $\frac{\tilde{N}_\mathrm{BS}!{\tilde{N}_\mathrm{UE}}^K}{N_\mathrm{RF}!(\tilde{N}_\mathrm{BS}-N_\mathrm{RF})!}$ feasible configurations \cite{interlaced_beam_training_allocation}. This observation motivates a sequential decision view of the original joint beam selection problem: the system can incrementally construct one joint configuration by selecting beams step by step.

Hence, we reformulate coordinated beam selection over the candidate subsets as a finite-horizon leader-follower Markov game \cite{decentralized_cooperative_rl}. This reformulation does not change the original ESE maximization objective in \eqref{mu_mimo_obj}. Instead, it provides a tractable decision model for constructing the joint beam selection solution sequentially. In each episode, the system constructs one joint configuration step by step. In this paper, we adopt the common assumption that the number of simultaneously served users equals the number of RF chains, i.e., $K=N_\mathrm{RF}$. The BS-side agent first selects a transmit beam because BS beams are shared across users and shape the effective multi-user channel, and the user-side agent then selects the receive beam conditioned on the previous BS-side decisions.

\begin{figure*}[t]
    \centering
    \includegraphics[width=0.99\linewidth]{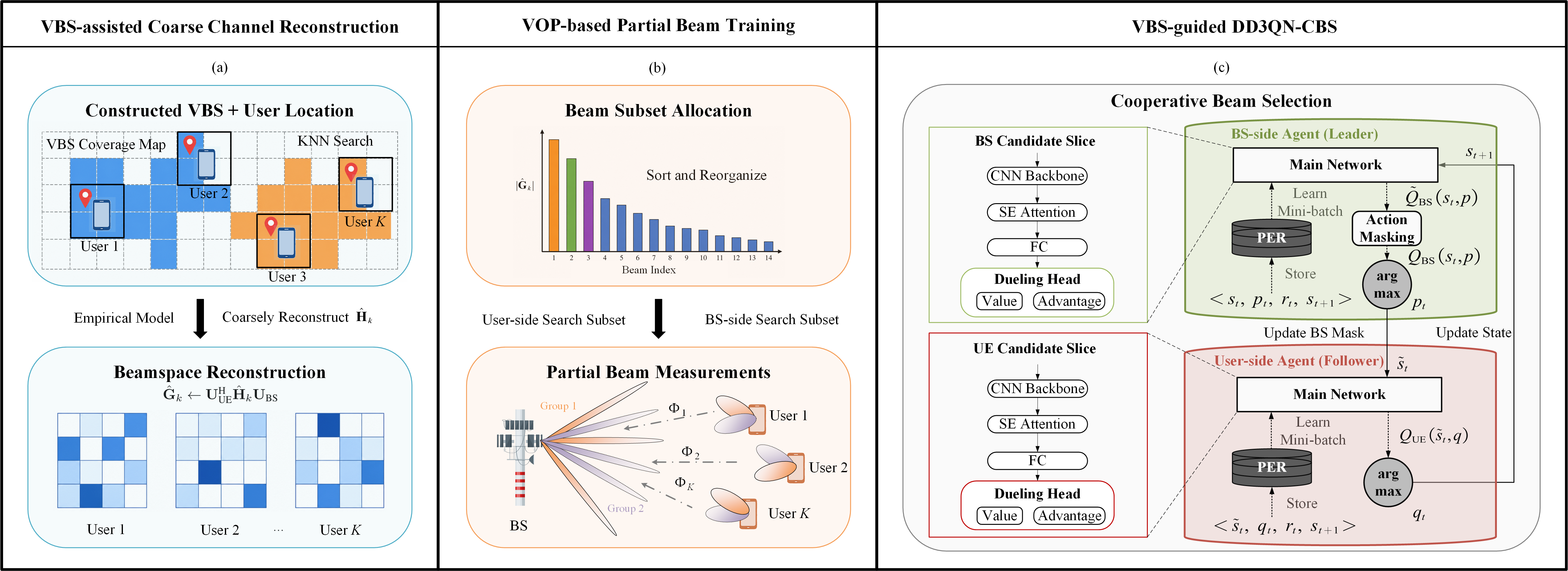}
    \caption{Proposed VBS-guided MU-MIMO beam management framework. (a) After constructing the VBS database offline, the BS performs coarse channel reconstruction to obtain the VBS-assisted coarse beamspace representation. (b) A VOP-based partial beam training scheme measures the selected candidate beam entries to refine this representation. (c) A dual-agent DD3QN-CBS algorithm performs interference-aware beam selection.}
    \label{fig:decision_auxiliary_beam}
\end{figure*}

Under this formulation, the state records both the beamspace features and the beam decisions already made. Since these selected beams are explicitly embedded into the state, the next state and immediate rewards depend only on the current state and the current actions $(p_t,q_t)$, making the sequential construction compatible with the Markov property. To further improve learning efficiency, we introduce BS-side action masking to avoid invalid reuse of already selected transmit beams and design dense intermediate rewards to alleviate the sparse terminal-reward issue. The overall framework is illustrated in Fig. \ref{fig:decision_auxiliary_beam}(c).

The resulting Markov game is formulated as
\begin{gather}
    \mathcal{M}=\left\langle \mathcal{S},\mathcal{A}_{\mathrm{BS}},\mathcal{A}_{\mathrm{UE}},\mathcal{P},\mathcal{R}_{\mathrm{BS}},\mathcal{R}_{\mathrm{UE}},\gamma\right\rangle,
\end{gather}
where $\mathcal{S}$ is the state space, $\mathcal{A}_{\mathrm{BS}}$ and $\mathcal{A}_{\mathrm{UE}}$ are the action spaces for the BS-side and user-side agents, respectively, $\mathcal{R}_{\mathrm{BS}}$ and $\mathcal{R}_{\mathrm{UE}}$ are the reward functions for the BS-side and user-side agents, and $\gamma$ is the discount factor. The transition kernel is defined as $s_{t+1}\sim \mathcal{P}(\cdot\mid s_t,p_t,q_t)$. 
We index the decision step as $t=0,\ldots,K-1$, while the physical user index is $k=1,\ldots,K$, thus step $t$ corresponds to the $(t+1)$-th scheduled user. The BS-side policy $\pi_{\mathrm{BS}}$ first selects a relative transmit-beam index $p_t$ according to $\pi_{\mathrm{BS}}(p_t\mid s_t)$. Conditioned on this leader action, the user-side policy $\pi_{\mathrm{UE}}$ selects a relative receive-beam index $q_t$ according to $\pi_{\mathrm{UE}}(q_t\mid \tilde{s}_t)$, where $\tilde{s}_t\triangleq(s_t,p_t)$ augments the state with the BS-side action \cite{decentralized_cooperative_rl}. The corresponding physical codeword indices are obtained from the candidate subsets as $\tilde{p}_t\triangleq[\mathcal{I}_\mathrm{BS}]_{p_t}$ and $\tilde{q}_t\triangleq[\mathcal{I}_\mathrm{UE}^{t+1}]_{q_t}$. This hierarchy keeps the BS-side decision visible to the user-side policy and therefore matches the interference-coupled nature of MU-MIMO beam selection.

Each agent $i\in\{\mathrm{BS},\mathrm{UE}\}$ receives an instantaneous reward $r_t^i$. Although the intermediate rewards are different, the two agents are cooperative because they share the same terminal objective, namely maximizing the ESE in \eqref{mu_mimo_obj}. The policy pair $\pi=(\pi_{\mathrm{BS}},\pi_{\mathrm{UE}})$ is therefore trained to maximize the expected discounted return
\begin{gather}
    J_i = \mathbb{E}_{\pi, \mathcal{P}}\left[\sum_{\ell=0}^{K-1-t} \gamma^\ell r_{t+\ell}^{i}\right].
\end{gather}
To evaluate this long-term return, each agent learns an action-value Q-function $ Q(\cdot) $, which quantifies the expected cumulative discounted reward of taking a specific action under state $s_t$ and subsequently following policy $\pi$.

Following the hierarchical information structure, the BS-side agent maintains $Q_{\mathrm{BS}}(s_t, p_t)$ to evaluate its transmit beam selections. Since the user-side agent observes the leader's action $p_t$ before executing its own, its decision-making is modeled by the joint action-value function $Q_{\mathrm{UE}}(s_t, p_t, q_t)$. These action-value functions satisfy the Bellman optimality equations, which can be expressed as
\begin{small}
\begin{gather}
\begin{aligned}
Q_{\mathrm{BS}}^\star(s_t, p_t)
&= \mathbb{E}_{\mathcal{P}}\!\left[
r_t^\mathrm{BS}
+ \gamma \max_{p_{t+1}}
Q_{\mathrm{BS}}^\star(s_{t+1}, p_{t+1}) \right], \\
Q_{\mathrm{UE}}^\star(s_t, p_t, q_t)
&= \mathbb{E}_{\mathcal{P}}\!\left[
r_t^\mathrm{UE}
+ \gamma \max_{q_{t+1}}
Q_{\mathrm{UE}}^\star(s_{t+1}, p_{t+1}^\star, q_{t+1}) \right],
\end{aligned}
\end{gather}
\end{small}
where $p_{t+1}^\star = \arg\max_p Q_{\mathrm{BS}}^\star(s_{t+1}, p)$ represents the predicted optimal action of the leader at the next state, and the expectation is taken over the environment transition kernel $\mathcal{P}$.

The above equations define the coupled optimality condition under ideal conditions. However, finding exact analytical solutions for $Q_{\mathrm{BS}}^\star$ and $Q_{\mathrm{UE}}^\star$ is computationally intractable due to the continuous and high-dimensional observation space. To circumvent this challenge, the proposed DD3QN-CBS framework employs dueling double deep-Q-network (D3QN) \cite{d3qn} to parameterize and approximate these optimal Q-functions, denoted as $Q_{\mathrm{BS}}(s_t, p_t; \bm{\Theta}_{\mathrm{BS}})$ and $Q_{\mathrm{UE}}(s_t, p_t, q_t; \bm{\Theta}_{\mathrm{UE}})$, where $\bm{\Theta}_{\mathrm{BS}}$ and $\bm{\Theta}_{\mathrm{UE}}$ represent the parameters of the online networks. 

To mitigate the inherent overestimation bias in discrete action spaces, the target values $y_t^{\mathrm{BS}}$ and $y_t^{\mathrm{UE}}$ are evaluated using the double Q-network principle by decoupling action selection from evaluation, which can be expressed as
\begin{gather}
\begin{aligned}
y_t^{\mathrm{BS}}
&= r_t^{\mathrm{BS}}
+ \gamma Q_{\mathrm{BS}}(s_{t+1}, p_{t+1}^\star; \bm{\Theta}_{\mathrm{BS}}'),\\
y_t^{\mathrm{UE}}
&= r_t^{\mathrm{UE}}
+ \gamma Q_{\mathrm{UE}}(s_{t+1}, p_{t+1}^\star, q_{t+1}^\star; \bm{\Theta}_{\mathrm{UE}}'),
\end{aligned}
\end{gather}
where $\bm{\Theta}'$ denotes the parameters of the target networks, and $ p_{t+1}^\star = \arg\max_p Q_{\mathrm{BS}}(s_{t+1}, p; \bm{\Theta}_{\mathrm{BS}})$ and $ q_{t+1}^\star = \arg\max_q Q_{\mathrm{UE}}(s_{t+1}, p_{t+1}^\star, q; \bm{\Theta}_{\mathrm{UE}})$ are the actions selected by the corresponding online networks. To enhance learning stability and sample efficiency under sparse feedback, both agents adopt a dueling network architecture that decomposes the parameterized action-value function into a state-value stream and an advantage stream as follows
\begin{equation}
    \begin{aligned}
        Q_i(s,a;\bm{\Theta}_i)
        &= V_i(s;\bm{\Theta}_i) + A_i(s,a;\bm{\Theta}_i) \\
        &\quad -\frac{1}{|\mathcal{A}_i|}
        \sum_{a'\in\mathcal{A}_i} A_i(s,a';\bm{\Theta}_i).
    \end{aligned}
    \label{eq:dueling_q}
\end{equation}

The above formulation establishes the mathematical principles governing how the two agents interact and how their parameterized Q-networks are updated. To complete the algorithmic framework, the subsequent sections elaborate on the concrete definitions of the foundational DRL components, i.e., the state space, action space, reward function, and network architecture.

\subsubsection{State Space}

The state space is structured to capture both the channel-quality information and the sequential decision history. Specifically, the BS-side and user-side agents share a composite input tensor $\mathfrak{G}_{\mathrm{in}}\in\mathbb{R}^{6\times\tilde{N}_{\mathrm{BS}}\times\tilde{N}_{\mathrm{UE}}\times K}$, constructed by stacking six information channels: the log-magnitude coarse beamspace features $\{\mathbf{X}_k^{\mathrm{coar}}\}_{k=1}^{K}$, the log-magnitude online beamspace features $\{\mathbf{X}_k^{\mathrm{meas}}\}_{k=1}^{K}$, the measurement mask $\mathfrak{M}$, the BS selection mask $\mathfrak{S}_t^{\mathrm{BS}}$, the user selection mask $\mathfrak{S}_t^{\mathrm{UE}}$, and the current-user mask $\mathfrak{C}_t$. The four masks are binary, where the measurement mask identifies entries obtained through VOP-based training, the selection masks record previously selected beams, and the current-user mask identifies the user undergoing beam selection.

The selection masks are updated according to the leader-follower order. After the BS-side agent selects beam index $p_t$ for the $(t+1)$-th scheduled user, the corresponding BS-side vector in the current user slice is marked as selected, which is expressed as
\begin{gather}
    [\mathfrak{S}_t^{\mathrm{BS}}]_{p_t,:,t+1}
    = \mathbf{1}_{\tilde{N}_{\mathrm{UE}}}.
\end{gather}
This partially updated state is then passed to the user-side agent as a conditional observation. After the user-side agent selects beam index $q_t$, the corresponding user-side vector is updated as
\begin{gather}
    [\mathfrak{S}_t^{\mathrm{UE}}]_{:,q_t,t+1}
    = \mathbf{1}_{\tilde{N}_{\mathrm{BS}}}.
\end{gather}
These updates make the current BS-side decision visible to the user-side agent and retain the accumulated beam selections across the episode. The BS selection history also identifies transmit beams occupied by previously served users. The current-user mask advances to the next scheduled user after each completed beam-pair selection, while the measurement mask remains fixed throughout the episode.

\subsubsection{Action Space}
To limit strong multi-user interference, the action space should be constrained against highly correlated beam assignments that cause equivalent channel rank degradation \cite{interlaced_beam_training_allocation,joint_beam_selection_precoding}. Specifically, if multiple concurrent users are assigned to the identical BS codeword, the resulting effective channel matrix will contain identical columns, rendering it structurally low-rank. Consequently, the composite channel-precoder matrix cannot be diagonalized regardless of the digital precoder design $\mathbf{F}_\mathrm{BB}$ at the BS-side, thereby weakening spatial multiplexing capability and reducing the ESE.

Instead of penalty-based collision avoidance, we em-
ploy action masking during decision making \cite{safe_action_masking,past_action_masking} to exclude beam collisions structurally. This maintains a consistent network architecture and accelerates training by allowing agents to focus exclusively on ranking the relative quality of valid beam pairs.

Following the same leader-follower order, the BS-side feasible action set is defined first. Starting from $\mathcal{A}_\mathrm{BS}^{0}=\{1,\ldots,\tilde{N}_\mathrm{BS}\}$, the feasible set is updated recursively by removing the relative beam index $p_{t-1}$ selected in the previous step, which is expressed as
\begin{gather}
    \mathcal{A}_\mathrm{BS}^{t}
    = \mathcal{A}_\mathrm{BS}^{t-1}\setminus\{p_{t-1}\},
    \quad t=1,\ldots,K-1.
\end{gather}
The masking process is implemented by forcing the Q-values of these previously selected BS-side beams to $-\infty$. Consequently, while the output dimension of the BS-side network remains fixed at $ |\mathcal{A}_\mathrm{BS}^0| $, its masked Q-value function is defined as
\begin{gather}
    \tilde{Q}_\mathrm{BS}(s_t, p) = \begin{cases}
    Q_\mathrm{BS}(s_t, p), & p \in \mathcal{A}_\mathrm{BS}^t \\
    -\infty, & \text{otherwise}
    \end{cases}.
\end{gather}
After the BS-side action is fixed, the user-side action space $\mathcal{A}_\mathrm{UE}^t$ represents the relative beam indices within the candidate subset of the currently scheduled user. Although the user-side network has a fixed discrete output dimension $\tilde{N}_\mathrm{UE}$, the decision at step $t$ is executed only for that specific user. This parameter-sharing design can be viewed as $K$ homogeneous user agents sharing a single Q-network, which improves sample efficiency and generalization without changing the underlying MU-MIMO coupling. 

For exploitation, the BS-side and user-side actions are selected sequentially according to
\begin{gather}
    p_t = \arg\max_{p\in \mathcal{A}_\mathrm{BS}^0} \tilde{Q}_\mathrm{BS}(s_t, p), \\
    q_t= \arg\max_{q\in \mathcal{A}_\mathrm{UE}^t} Q_\mathrm{UE}(\tilde{s}_t, q).
\end{gather}
During training, an $\epsilon$-greedy exploration strategy is adopted, where each agent randomly samples from its feasible action set with probability $\epsilon$ and follows the greedy action above with probability $1-\epsilon$.
The resulting joint action at time step $t$ is denoted as $(p_t, q_t)$. To further enhance sample efficiency, we employ the prioritized experience replay (PER) buffer \cite{PER}.

Once the feasible action space is defined, the next challenge is to provide informative training signals before the final ESE becomes available.

\subsubsection{Reward Function} 
The reward design follows the same principle: the terminal objective should remain the ESE, but the learning signal should not be delayed until the end of the episode. Otherwise, the learning process would lead to inefficient credit assignment and slow convergence. Therefore, the BS-side and user-side rewards are defined in the same decision order using the estimated SINR proxies, while the terminal feedback remains the actual ESE.

Following the leader-follower decision structure, the BS-side immediate reward design should evaluate whether the current transmit beam provides useful multi-user coverage while avoiding overlap with previously selected BS beams. For notation clarity, let $\mathcal{B}_{<t}\triangleq\{\tilde{p}_i\}_{i=0}^{t-1}$ denote the set of BS beams selected before step $t$, with $\mathcal{B}_{<0}=\emptyset$. \textcolor{black}{Here, $\mathbf{G}_k$ denotes the ground-truth beamspace for user $k$, used only for reward calculation during offline training. Online decisions use the observation defined above.} The corresponding utility function is defined as
\begin{align}
    U_t^\mathrm{BS}(p_t)
    &=
    \sum_{k=1}^{K}
    \frac{
    P_\mathrm{T}\max\limits_{\tilde{q}\in\mathcal{I}_\mathrm{UE}^{k}}
    |[\mathbf{G}_{k}]_{\tilde{q},\tilde{p}_t}|^2
    }{
    P_\mathrm{T}\sum\limits_{\tilde{p}\in\mathcal{B}_{<t}}
    \max\limits_{\tilde{q}\in\mathcal{I}_\mathrm{UE}^{k}}
    |[\mathbf{G}_{k}]_{\tilde{q},\tilde{p}}|^2
    +N_0W
    }.
    \label{bs_intermediate_utility}
\end{align}
The user-side immediate reward is conditioned on the BS-side action and should evaluate the selected receive beam quality while considering SINR. Thus, we also define the user-side utility function as
\begin{align}
    U_t^\mathrm{UE}(p_t,q_t)
    &=
    \frac{
    P_\mathrm{T}|[\mathbf{G}_{t+1}]_{\tilde{q}_t,\tilde{p}_t}|^2
    }{
    P_\mathrm{T}\sum\limits_{\tilde{p}\in\mathcal{B}_{<t}}
    |[\mathbf{G}_{t+1}]_{\tilde{q}_t,\tilde{p}}|^2
    +N_0W
    }.
    \label{ue_intermediate_utility}
\end{align}
The resulting reward functions are formulated as
\begin{align}
    r_t^\mathrm{BS} = 
    \begin{cases}
        \alpha_\mathrm{BS} g\big[U_t^\mathrm{BS}(p_t)\big], & t < K-1, \\[1ex]
        \mathrm{ESE}, & t = K-1,
    \end{cases}
\end{align}
\begin{align}
    r_t^\mathrm{UE} = 
    \begin{cases}
        \alpha_\mathrm{UE} g\big[U_t^\mathrm{UE}(p_t,q_t)\big], & t < K-1, \\[1ex]
        \mathrm{ESE}, & t = K-1,
    \end{cases}
\end{align}
where $ g[\cdot] $ is a normalization function that scales the reward to $ [0,1] $, and $ \alpha_\mathrm{BS}, \alpha_\mathrm{UE} $ are scaling factors. 

\textcolor{black}{During deployment, the agents observe the VBS-derived beamspace prior, VOP-measured entries, and their associated masks. Differences between the simulated training environment and the deployment environment can alter these observations and the resulting policy behavior. Measurements and communication feedback collected in the target environment can inform subsequent policy adaptation when the distribution shift is substantial.}

Upon completing the sequential decisions, the selected BS beam set $\mathcal{B}_{<K} \subset \mathcal{I}_{\mathrm{BS}}$ and the mapped user-side beam indices determine the analog beamforming matrices. Specifically, the selected DFT submatrices for the analog precoder at the BS and the analog combiner for the $k$-th user are represented as $\mathbf{F}_{\mathrm{RF}} = [\mathbf{U}_\mathrm{BS}]_{:,\mathcal{B}_{<K}}$ and $\mathbf{w}_{k} = [\mathbf{U}_{\mathrm{UE}}]_{:,\tilde{q}_{k-1}}$ for $k=1,\ldots,K$, respectively. \textcolor{black}{After the analog beamformer and user combiners are selected, the resulting low-dimensional effective channel is obtained for MMSE digital precoding.} Consequently, the effective channel vector of the $k$-th user is determined by $\mathbf{h}_{\mathrm{eff}}^k = \mathbf{w}_k^{\mathsf{H}} \mathbf{H}_k \mathbf{F}_{\mathrm{RF}} \in \mathbb{C}^{1 \times N_{\mathrm{RF}}}$. By stacking these individual vectors, the composite effective channel matrix after analog beam selection is formed as $\mathbf{H}_{\mathrm{eff}} = [\mathbf{h}_{\mathrm{eff}}^1; \ldots; \mathbf{h}_{\mathrm{eff}}^K] \in \mathbb{C}^{K \times N_{\mathrm{RF}}}$. 

Utilizing this low-dimensional effective channel matrix $\mathbf{H}_{\mathrm{eff}}$, the system performs baseband digital precoding. Considering both computational complexity and multi-user interference mitigation, the minimum mean-square error (MMSE) precoder is adopted for digital baseband precoding, which can be obtained by \cite{naive_mmse}
\begin{equation}
\mathbf{F}_{\mathrm{BB}} = \bigl(\mathbf{H}_{\mathrm{eff}}^{\mathsf{H}}\mathbf{H}_{\mathrm{eff}} + \eta \mathbf{I}\bigr)^{-1} \mathbf{H}_{\mathrm{eff}}^{\mathsf{H}},
\end{equation}
where the regularization parameter $\eta$ is set to $KN_0W/P_{\mathrm{T}}$ to balance noise enhancement and inter-user interference suppression. 
The resulting digital precoder is normalized to satisfy the total transmit power constraint in \eqref{mu_mimo_cons3}.

\subsubsection{Network Architecture}

\textcolor{black}{The beamspace observations are organized as regular tensors indexed by BS beam candidates, user beam candidates, and scheduled users. This structure motivates a candidate-wise convolutional neural network (CNN) for extracting beamspace features, together with a lightweight squeeze-and-excitation (SE) module \cite{sequeeze_excitation} for adaptively weighting feature channels. Multi-user coordination is supported by the shared observations and sequential BS-side and user-side decisions, with interference-aware rewards guiding the training process. Although graph neural networks can explicitly model inter-user relationships, the proposed architecture directly exploits the structured beamspace observations.}

\textcolor{black}{The BS-side and user-side agents employ independently parameterized dueling Q-networks, each using a shared CNN encoder across its candidate actions. For the BS-side agent, each candidate beam is represented by an observation slice $[\mathfrak{G}_{\mathrm{in}}]_{:,b,:,:}\in\mathbb{R}^{6\times\tilde{N}_{\mathrm{UE}}\times K}$. Similarly, the user-side agent processes $[\mathfrak{G}_{\mathrm{in}}]_{:,:,u,:}\in\mathbb{R}^{6\times\tilde{N}_{\mathrm{BS}}\times K}$ for each receive-beam candidate, using the updated observation conditioned on the BS-side decision.}

\textcolor{black}{Each candidate slice passes through two $3\times3$ convolutional layers with channel dimensions $6\rightarrow16\rightarrow16$, each followed by GroupNorm and LeakyReLU. The SE module follows the second convolution and uses a reduction ratio of four with excitation dimensions $16\rightarrow4\rightarrow16$. Its final excitation layer is zero-initialized, and the channel weights are obtained using twice the sigmoid activation, yielding an identity mapping at initialization. Adaptive average pooling, a fully connected (FC) projection, and LayerNorm then produce a 128-dimensional embedding for each candidate.}

\textcolor{black}{Mean and element-wise maximum pooling across candidate embeddings form a 256-dimensional shared context. The advantage stream combines each 128-dimensional candidate embedding with this context and a 128-dimensional element-wise interaction between the candidate embedding and the mean context, resulting in a 512-dimensional input. The value stream uses the shared context directly. The advantage and value streams employ FC layers with dimensions $512\rightarrow128\rightarrow1$ and $256\rightarrow128\rightarrow1$, respectively, and their outputs are combined through \eqref{eq:dueling_q}. }

\section{Simulation Results}
\label{simulation_results}
\begin{table}
    \footnotesize
    \renewcommand{\arraystretch}{1.}
    \centering
    \caption{Simulation parameters.}
    \label{setup_config}
    \begin{tabular}{cc}
        \toprule
        Parameter & Value \\
        \midrule
        Carrier Frequency $ f_\mathrm{c} $ & 40 GHz \\
        \hline
        Bandwidth $ W $ & 100 MHz \\
        \hline
        BS Position $ \mathbf{o}_\mathrm{BS} $ & [140, 60, 4] m \\
        \hline
        BS/user antenna size $ N_\mathrm{BS}, N_\mathrm{UE} $ & 128, 8 \\
        \hline
        Transmit Power $ P_\mathrm{T} $ & 40 dBm \\
        \hline
        \textcolor{black}{Effective Reflection Loss $\Gamma$ [dB]} & \textcolor{black}{10} \\
        \hline
        KNN for $ \mathcal{K}_0 $ and $ \mathcal{K}_v $ & 6, 3\\
        \hline
        VBS Grid Number $ D_\mathrm{x}, D_\mathrm{y} $ & 40, 40\\
        \hline
        Learning Rate & $ 10^{-4} $ \\
        \hline
        Discount Factor $ \gamma $ & 0.99 \\
        \hline
        Scaling Factor $ \alpha_\mathrm{BS}, \alpha_\mathrm{UE} $ & 2.5, 1 \\
        \hline
        Batch Size & 32 \\
        \hline
        Training episodes & 8400 \\
        \bottomrule
    \end{tabular}
\end{table}
\subsection{Simulation Setup}
The simulation setup is designed to verify whether the proposed offline VBS construction and online beam management pipeline remain effective in a realistic urban layout. The simulation framework is built upon a high-fidelity digital twin integrating Blender and Sionna \cite{sionna}. We reconstruct a realistic urban scenario from an OpenStreetMap (OSM) source extent in Mexia, Texas, spanning approximately $369$ m in the east--west direction and $234$ m in the north--south direction (Latitude: $31.6802^\circ$--$31.6823^\circ$N, Longitude: $-96.4826^\circ$--$-96.4787^\circ$W). Blender is employed to generate synthetic 3D LiDAR point clouds, while Sionna performs ray-tracing (RT) based channel modeling to capture realistic propagation characteristics. To emulate practical sensing imperfections, we introduce Gaussian measurement noise with a standard deviation of 10 cm and a random point drop rate of 10\% into the LiDAR data. Users are randomly distributed within unoccupied regions of the simulated scene. Detailed system parameters are summarized in Table \ref{setup_config}.

\subsection{Performance of Coarse Channel Reconstruction}

\begin{figure}[t]
    \centering
    \begin{subfigure}[b]{0.99\linewidth}
        \centering
        \includegraphics[width=\linewidth]{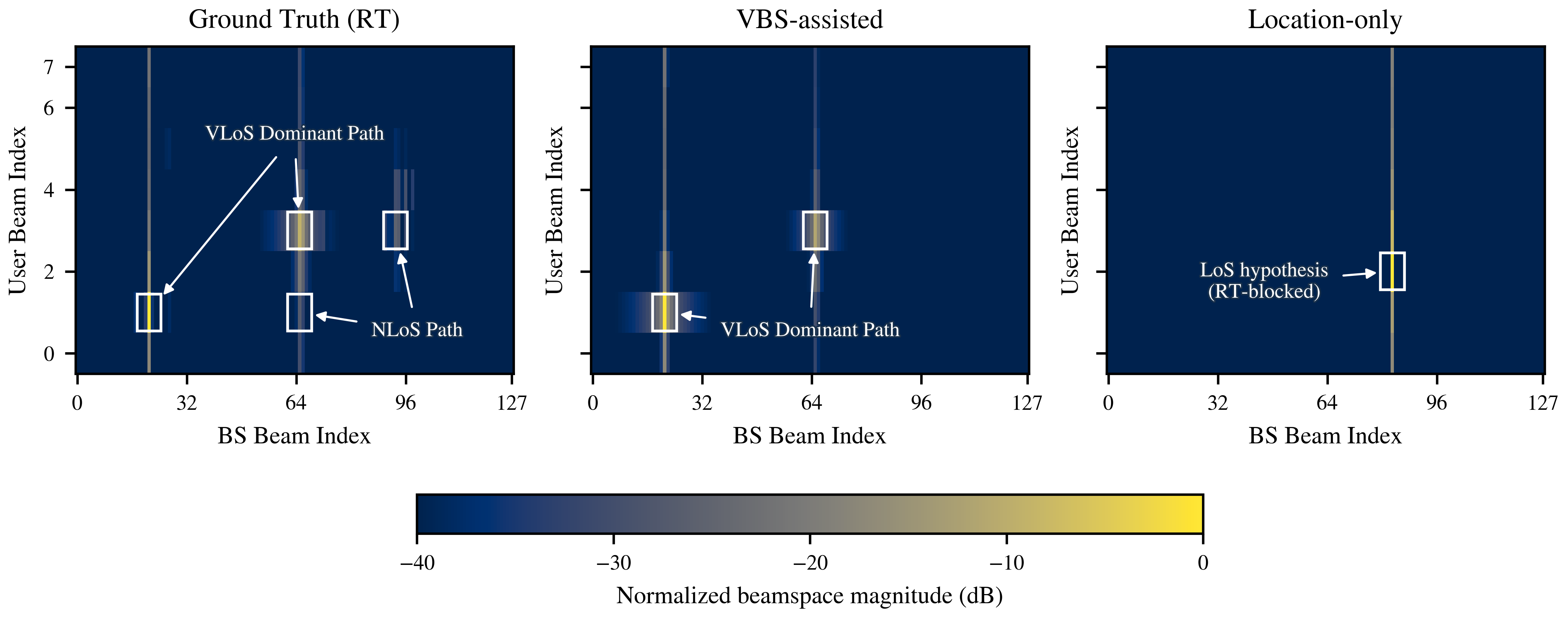}
        \caption{\textcolor{black}{Normalized beamspace comparison under blocked-LoS propagation.}}
        \label{fig:beamspace_triplet_case1}
    \end{subfigure}
    \begin{subfigure}[b]{0.99\linewidth}
        \centering
        \includegraphics[width=\linewidth]{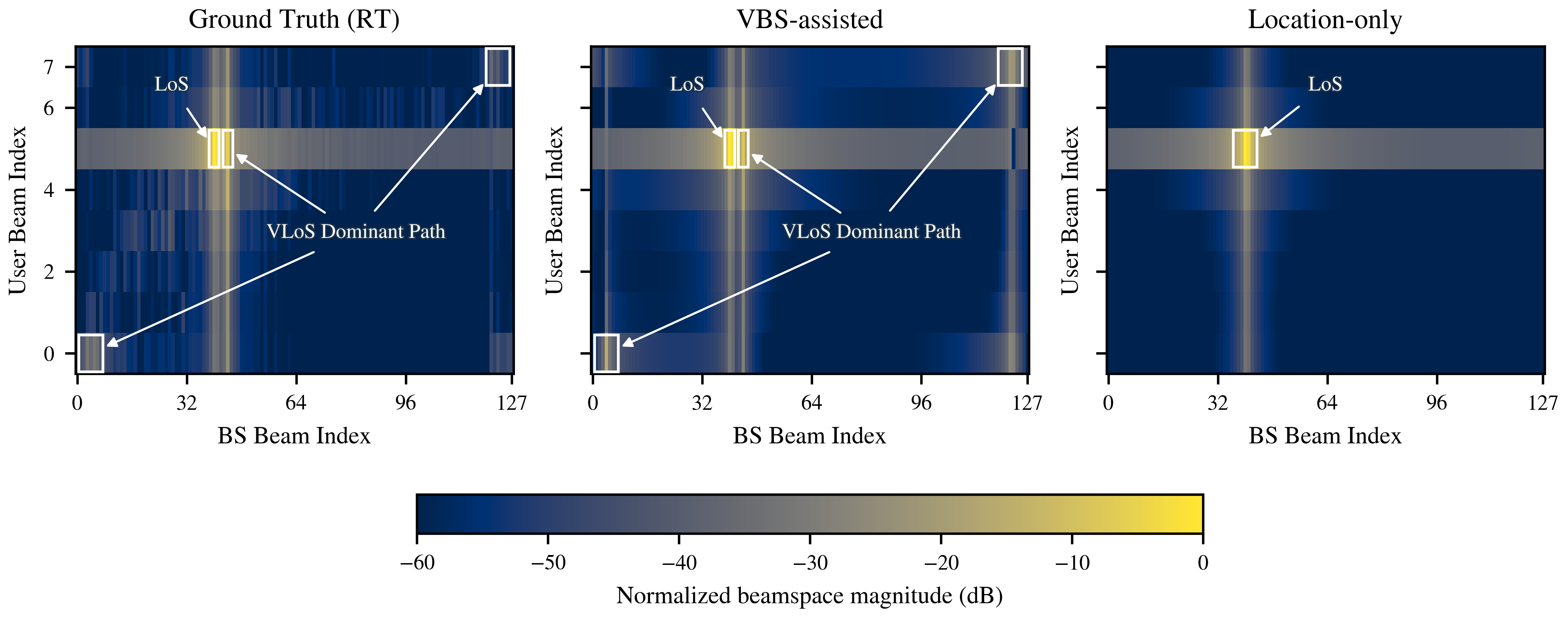}
        \caption{\textcolor{black}{Normalized beamspace comparison under mixed LoS/VLoS propagation.}}
        \label{fig:beamspace_triplet_case2}
    \end{subfigure}
    \caption{\textcolor{black}{Normalized beamspace magnitude comparison among VBS-assisted coarse reconstruction, location-only geometry, and the ground-truth RT beamspace.}}
    \label{fig:visualization_of_beamspace}
\end{figure}

The coarse channel reconstruction stage is evaluated by comparing the spatial magnitude distributions of the VBS-assisted coarse beamspace and the location-only geometric beamspace with the ground-truth RT beamspace.

\textcolor{black}{Fig.~\ref{fig:visualization_of_beamspace} compares the spatial beamspace magnitude distributions under two representative propagation conditions. In the blocked-LoS scenario shown in Fig.~\ref{fig:beamspace_triplet_case1}, the VBS-assisted reconstruction captures reflected angular components that are absent from the location-only representation. When LoS and reflected paths coexist, as shown in Fig.~\ref{fig:beamspace_triplet_case2}, both geometric representations identify the dominant LoS direction. The heatmaps are individually normalized by their maximum magnitudes to illustrate the relative spatial patterns.}

\textcolor{black}{To quantify the reconstruction accuracy over the service region, we compare the magnitude distributions of the reconstructed and reference beamspace matrices after independently normalizing them by their Frobenius norms. Let $\mathbf{G}_g$ and $\hat{\mathbf{G}}_g$ denote the reference and reconstructed beamspace matrices at service-grid location $g$, respectively. The magnitude NMSE is defined as}\begin{equation}
\textcolor{black}{\mathrm{NMSE}_{\mathrm{mag}}=10\log_{10}\!\left(\frac{1}{N_{\mathrm{g}}}\sum_{g=1}^{N_{\mathrm{g}}}\left\|\frac{|\mathbf{G}_g|}{\|\mathbf{G}_g\|_{\mathrm{F}}}-\frac{|\hat{\mathbf{G}}_g|}{\|\hat{\mathbf{G}}_g\|_{\mathrm{F}}}\right\|_{\mathrm{F}}^2\right),}
\label{eq:magnitude_nmse}
\end{equation}
\textcolor{black}{where $N_{\mathrm{g}}$ denotes the number of evaluated service-grid locations. Across 5,706 locations, the VBS-assisted reconstruction achieves a magnitude NMSE of $-1.523$ dB, compared with $-0.164$ dB for location-only geometry. The lower reconstruction error indicates that the VBS representation better preserves the relative spatial magnitude distribution of the reference RT beamspace. The VBS reconstruction relies on dominant first-order
reflections, while the full RT channels also contain
higher-order and diffuse propagation components.
To examine the angular information retained by this
approximation, we compare the modeled LoS and first-order
reflected paths with the full RT channels. Across the same
5,706 service-grid locations, these paths preserve the
strongest full-RT beam pair in 86.5\% of cases and achieve
an average Top-10 recall of 81.0\%. The retained dominant
beam directions provide a basis for the subsequent
VBS-guided candidate construction.}

\subsection{Performance of VOP-based Partial Beam Training}

\textcolor{black}{We first evaluate the candidate subsets constructed by Algorithm~\ref{beam_training_subset_selection} using the full-RT beamspace as reference. The evaluation includes 5,600 multi-user samples with 20 users per sample. With 64 BS-beam candidates and four receive-beam candidates per user, the selected subsets retain $25\%$ of the complete beam-pair space. The VBS-guided candidates achieve strongest-beam coverage of $87.45\%$, compared with $69.93\%$ for the geometry-based location-only baseline under the same candidate budget. }

\begin{table}[t]
    \centering
    \caption{\textcolor{black}{Averaged NMSE (dB) of VOP-based partial beam training under different measurement budgets.}}
    \label{beam_training_nmse}
    \begingroup
    \scriptsize
    \setlength{\tabcolsep}{6pt}
    \begin{tabular}{ccccc}
        \toprule
        $\bar{N}_{\mathrm{UE}}$
        & \multicolumn{4}{c}{$\bar{N}_{\mathrm{BS}}$} \\
        \cmidrule(lr){2-5}
        & 14 & 28 & 42 & 56 \\
        \midrule
        1
        & \textcolor{black}{$-1.83$}
        & \textcolor{black}{$-2.43$}
        & \textcolor{black}{$-2.79$}
        & \textcolor{black}{$-3.02$} \\
        2
        & \textcolor{black}{$-2.39$}
        & \textcolor{black}{$-3.45$}
        & \textcolor{black}{$-4.24$}
        & \textcolor{black}{$-4.83$} \\
        3
        & \textcolor{black}{$-2.59$}
        & \textcolor{black}{$-3.87$}
        & \textcolor{black}{$-4.91$}
        & \textcolor{black}{$-5.76$} \\
        \bottomrule
    \end{tabular}
    \endgroup
\end{table}

\textcolor{black}{Table~\ref{beam_training_nmse} evaluates the complex beamspace approximation obtained through VOP-based partial measurements under different measurement budgets. The evaluation uses 5,600 channel samples with 14 scheduled users per sample, following the antenna configuration in Table~\ref{setup_config}. Under the noiseless orthogonal-pilot protocol, the measured beamspace $\tilde{\mathbf{G}}_{s,k}$ contains the corresponding complex channel coefficients at measured beam pairs, while unmeasured entries are set to zero. Its NMSE relative to the reference beamspace $\mathbf{G}_{s,k}$ is defined as}
\textcolor{black}{\begin{equation}
\mathrm{NMSE}
=10\log_{10}\!\Bigg(
\frac{1}{SK}
\sum_{s=1}^{S}
\sum_{k=1}^{K}
\frac{
\|\mathbf{G}_{s,k}
-\tilde{\mathbf{G}}_{s,k}\|_{\mathrm{F}}^2
}{\|\mathbf{G}_{s,k}\|_{\mathrm{F}}^2
}\Bigg),
\label{eq:vop_nmse}
\end{equation}
where $S$ and $K$ denote the numbers of evaluated channel samples and scheduled users, respectively. Unlike the magnitude NMSE in \eqref{eq:magnitude_nmse}, this metric evaluates the complex beamspace coefficients without separately normalizing the measured beamspace matrix.}

\textcolor{black}{As shown in Table~\ref{beam_training_nmse}, the NMSE decreases from $-1.83$ to $-5.76$ dB as the BS- and receive-beam measurement budgets increase. Both measurement dimensions contribute to improved beamspace approximation, while the incremental gains from additional BS-beam measurements diminish for a fixed receive-beam budget. Thus, 
for the learning-based comparisons in the following subsection, the initial measurement budget is fixed at $\bar{N}_{\mathrm{BS}}=K$ and $\bar{N}_{\mathrm{UE}}=2$. The same budget is maintained across the VBS-guided candidate configurations.}

\textcolor{black}{We further evaluate the sensitivity of VBS-guided candidate selection to the effective reflection loss $\Gamma$ using the same 5,600 multi-user samples with 20 users per sample. Varying $\Gamma$ from 5 to 15 dB yields strongest-beam coverage between $85.87\%$ and $88.01\%$, compared with $87.45\%$ at the nominal value of 10 dB. The limited variation indicates that the dominant candidate directions remain largely consistent over the examined reflection-loss range.}

\textcolor{black}{To examine sensing--channel mismatch, we introduce geometric deviations, reflector-dependent attenuation, and incomplete reflector information into the VBS-derived prior while keeping the reference RT channels unchanged. Under the same $64\times4$ candidate budget and evaluation samples, a 0.5-m geometric perturbation reduces the strongest-beam coverage from $87.45\%$ to $85.56\%$. The coverage remains $80.55\%$ under combined mismatch, compared with $69.93\%$ for location-only geometry. These results show that VBS-guided candidate selection retains useful beam directions under the evaluated uncertainties, although incomplete reflector information can reduce candidate coverage.}

\subsection{Performance of DD3QN-CBS}
\begin{figure}
    \centering
    \begin{subfigure}[b]{0.48\linewidth}
        \centering
        \includegraphics[width=\linewidth]{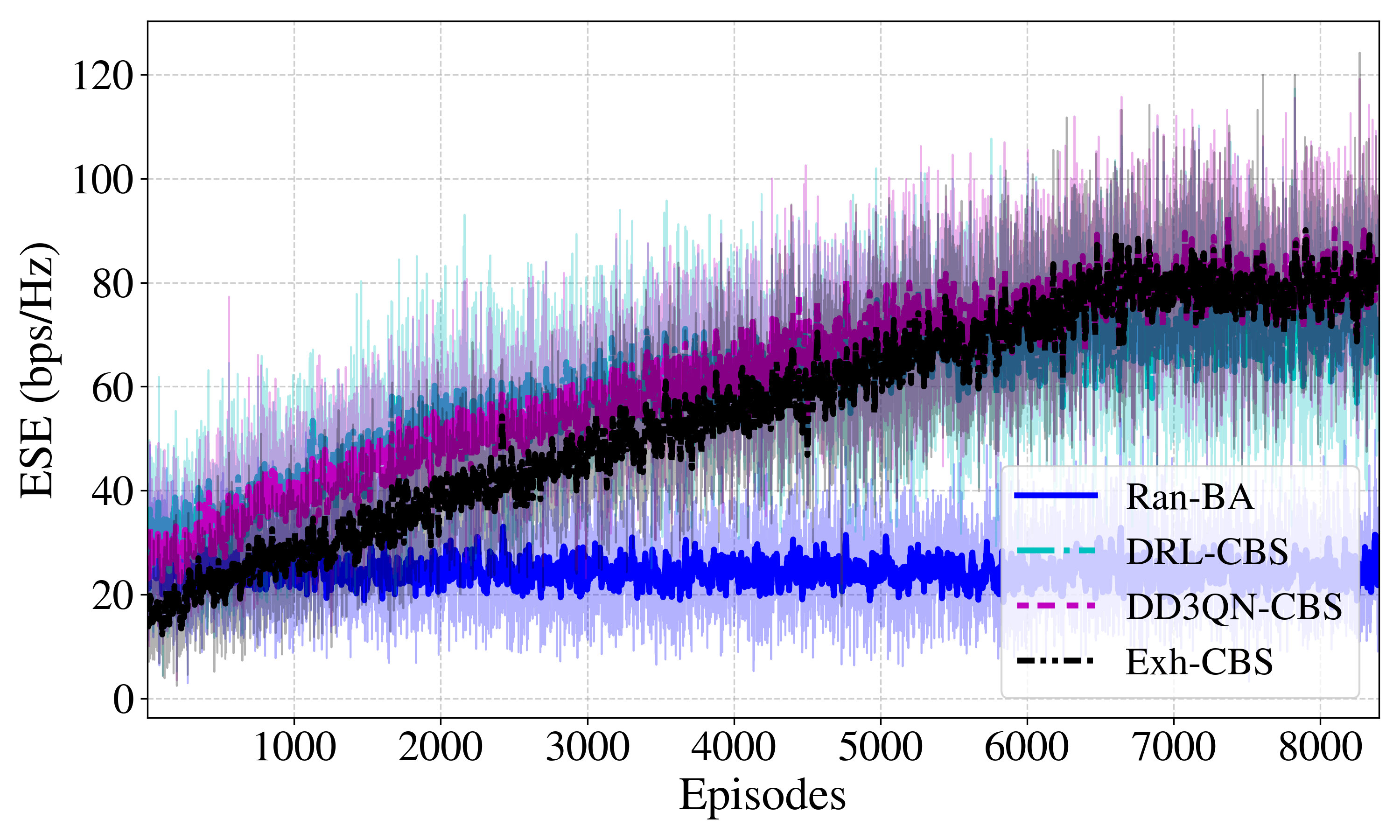}
        \caption{$N_\mathrm{RF}=14$.}
        \label{fig:train_se_d3qn_vs_baseline_params_10_14_64_-174}
    \end{subfigure}
    \begin{subfigure}[b]{0.48\linewidth}
        \centering
        \includegraphics[width=\linewidth]{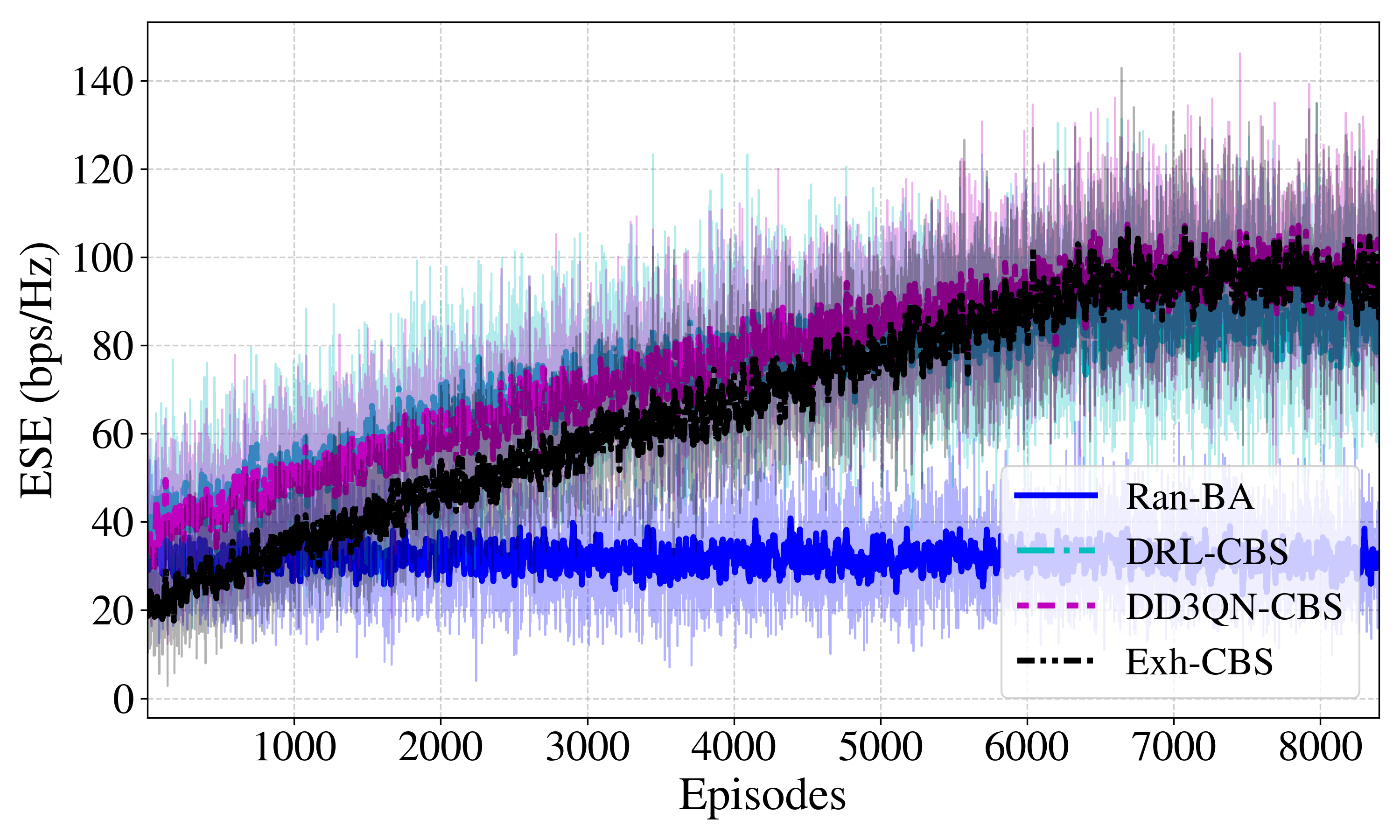}
        \caption{$N_\mathrm{RF}=18$.}
        \label{fig:train_se_d3qn_vs_baseline_params_10_18_64_-174}
    \end{subfigure}
    \begin{subfigure}[b]{0.48\linewidth}
        \centering
        \includegraphics[width=\linewidth]{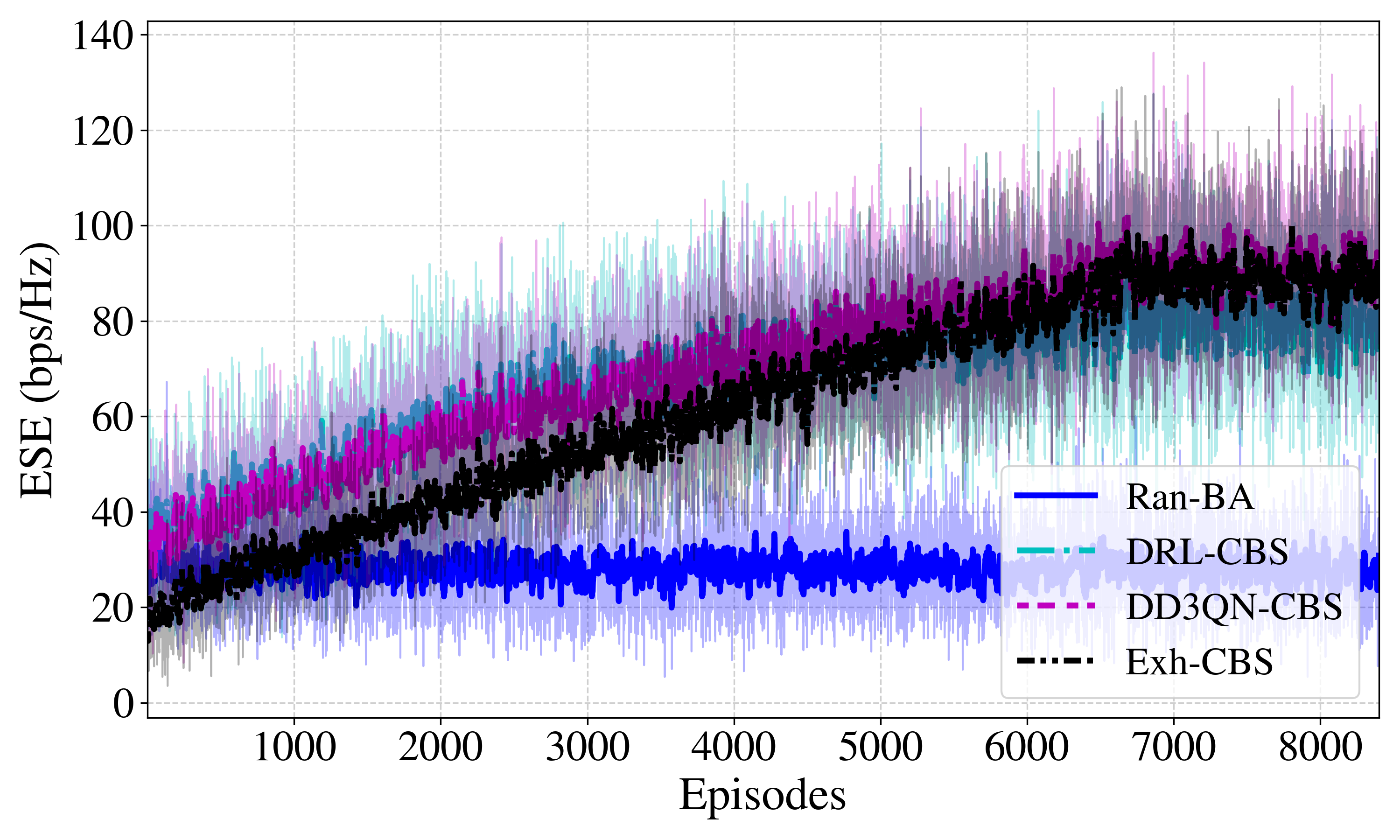}
        \caption{$N_0=-170$ dBm/Hz.}
        \label{fig:train_se_d3qn_vs_baseline_params_10_20_64_-170}
    \end{subfigure}
    \begin{subfigure}[b]{0.48\linewidth}
        \centering
        \includegraphics[width=\linewidth]{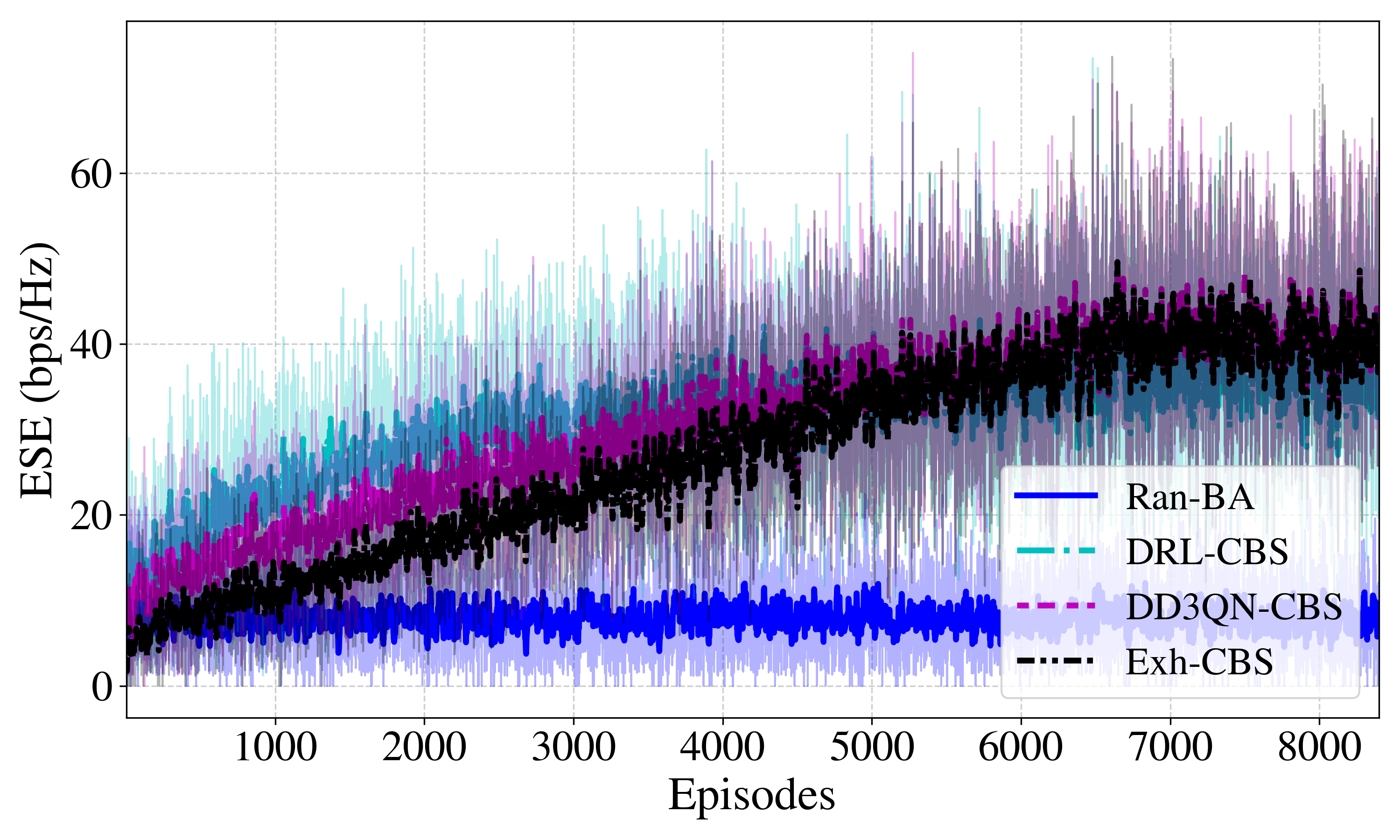}
        \caption{$N_0=-154$ dBm/Hz.}
        \label{fig:train_se_d3qn_vs_baseline_params_10_20_64_-154}
    \end{subfigure}
    \begin{subfigure}[b]{0.48\linewidth}
        \centering
        \includegraphics[width=\linewidth]{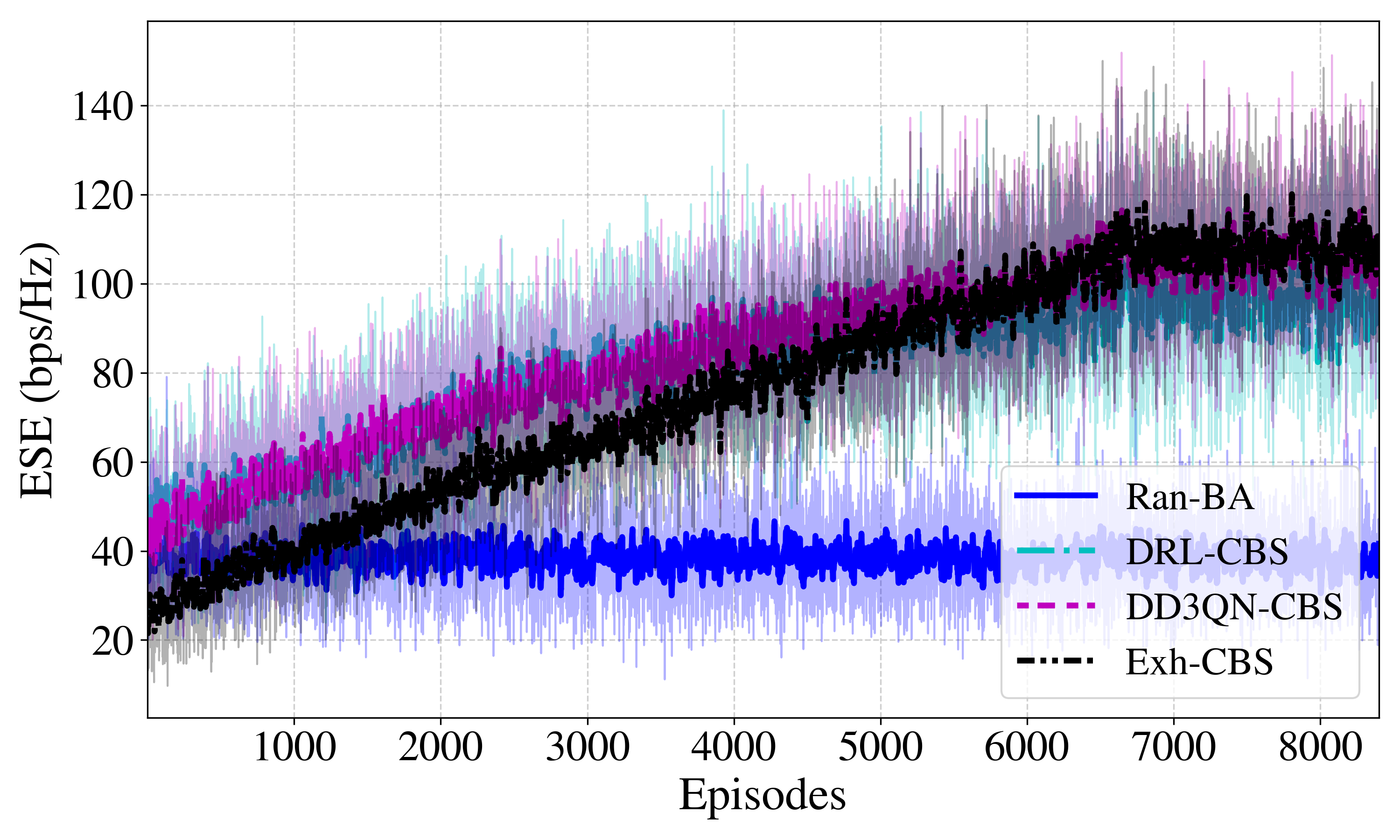}
        \caption{$\mathrm{SINR}_\mathrm{thres}=5$ dB.}
        \label{fig:train_se_d3qn_vs_baseline_params_5_20_64_-174}
    \end{subfigure}
    \begin{subfigure}[b]{0.48\linewidth}
        \centering
        \includegraphics[width=\linewidth]{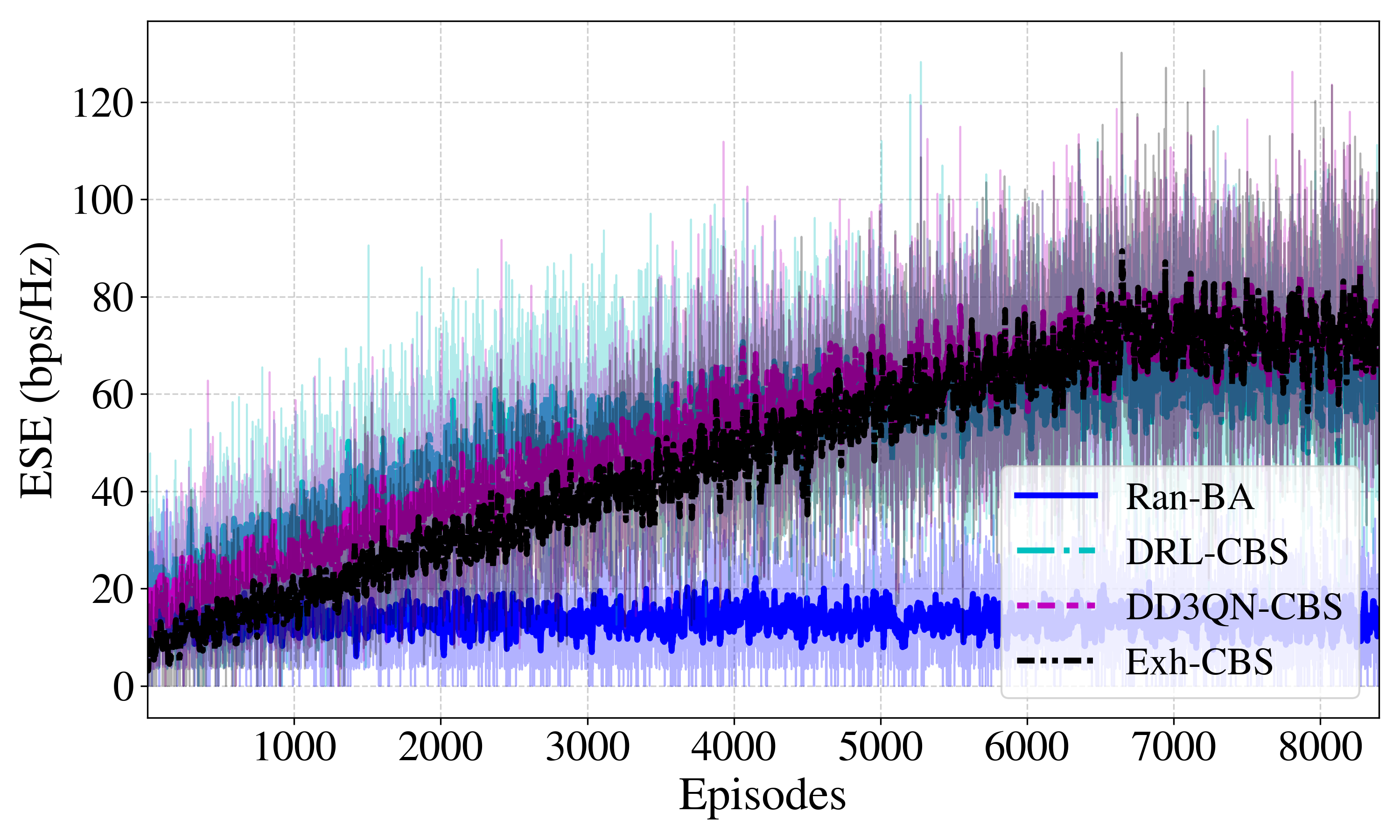}
        \caption{$\mathrm{SINR}_\mathrm{thres}=30$ dB.}
        \label{fig:train_se_d3qn_vs_baseline_params_30_20_64_-174}
    \end{subfigure}
    \caption{\textcolor{black}{Training ESE of the learning-based schemes under representative operating conditions.}}
    \label{beam_training_ese_comparison}
\end{figure}

\begin{figure}
    \centering
    \begin{subfigure}[b]{0.9\columnwidth}
        \centering
        \includegraphics[width=\linewidth]{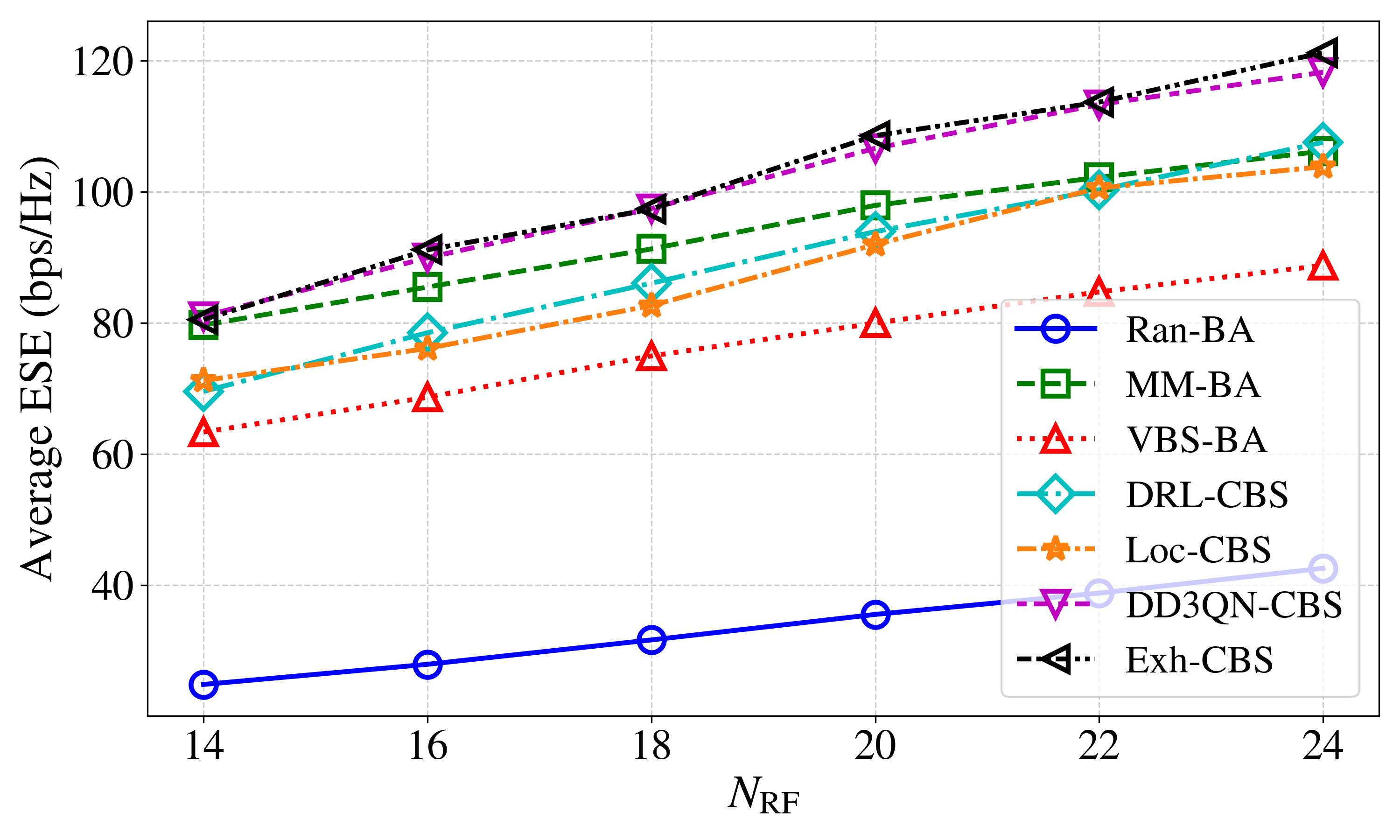}
        \caption{ESE versus different $N_\mathrm{RF}$ configurations.}
        \label{fig:N_rf_average_ESE}
    \end{subfigure}
    \begin{subfigure}[b]{0.9\columnwidth}
        \centering
        \includegraphics[width=\linewidth]{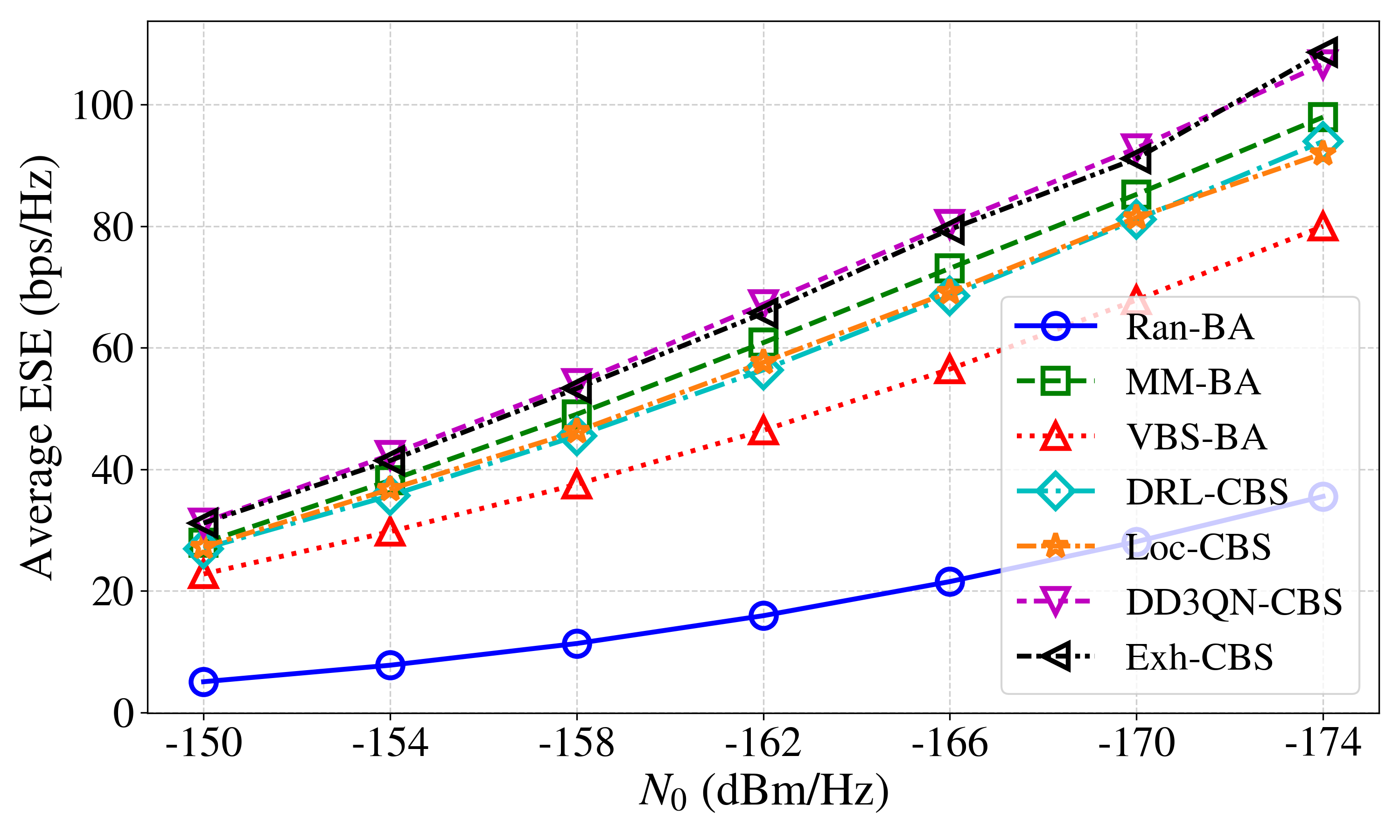}
        \caption{ESE versus different $N_0$ configurations.}
        \label{fig:Pnoise_average_ESE}
    \end{subfigure}
    \begin{subfigure}[b]{0.9\columnwidth}
        \centering
        \includegraphics[width=\linewidth]{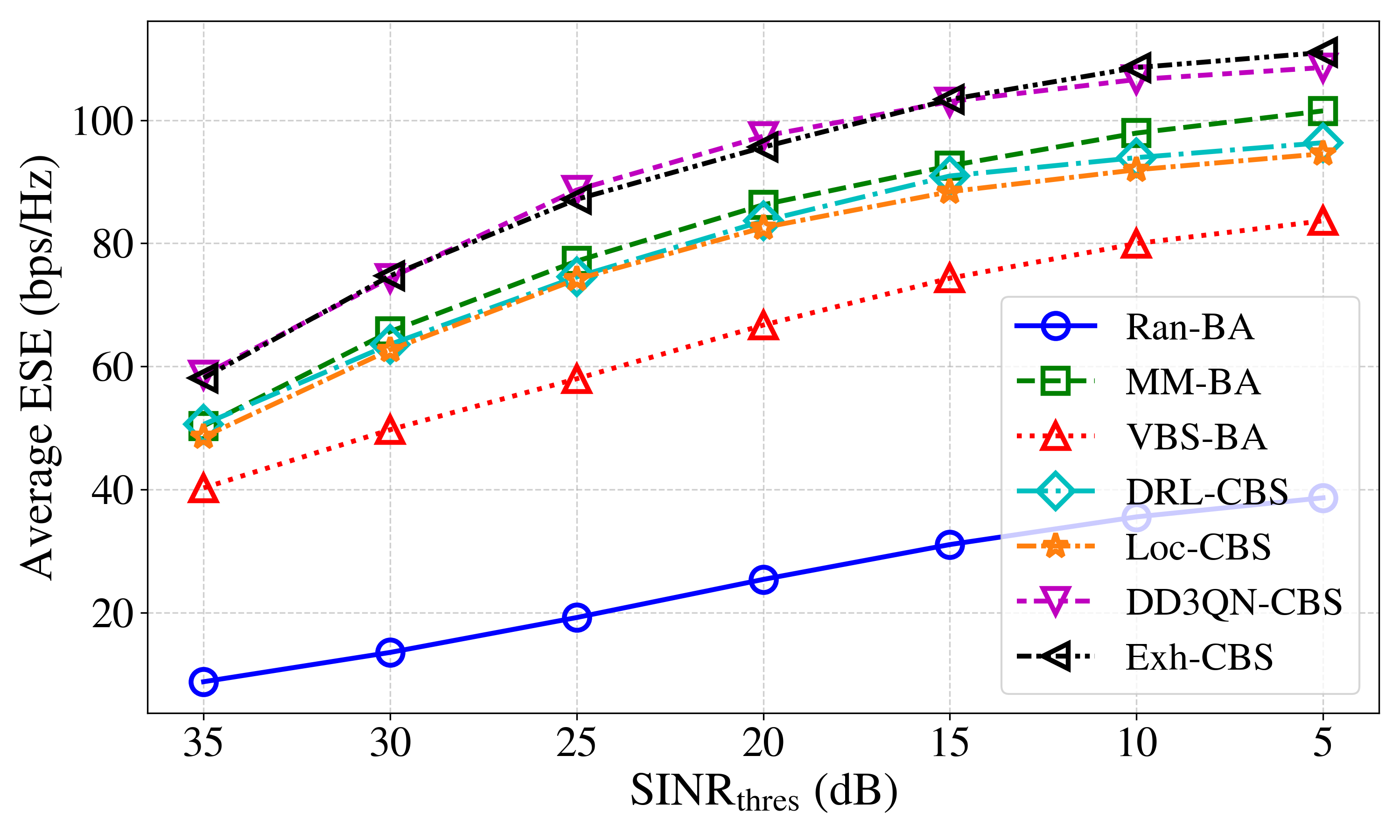}
        \caption{ESE versus different $\mathrm{SINR}_\mathrm{thres}$ configurations.}
        \label{fig:SINR_average_ESE}
    \end{subfigure}
    \caption{\textcolor{black}{Average ESE comparison under different system configurations. The default setting is $N_\mathrm{RF}=K=20$, $N_0=-174$ dBm/Hz, and $\mathrm{SINR}_\mathrm{thres}=10$ dB.}}
    \label{fig:average_ESE_comparison}
\end{figure}

\begin{figure}[t]
    \centering
    \begin{subfigure}[b]{0.48\linewidth}
        \centering
        \includegraphics[width=\linewidth]{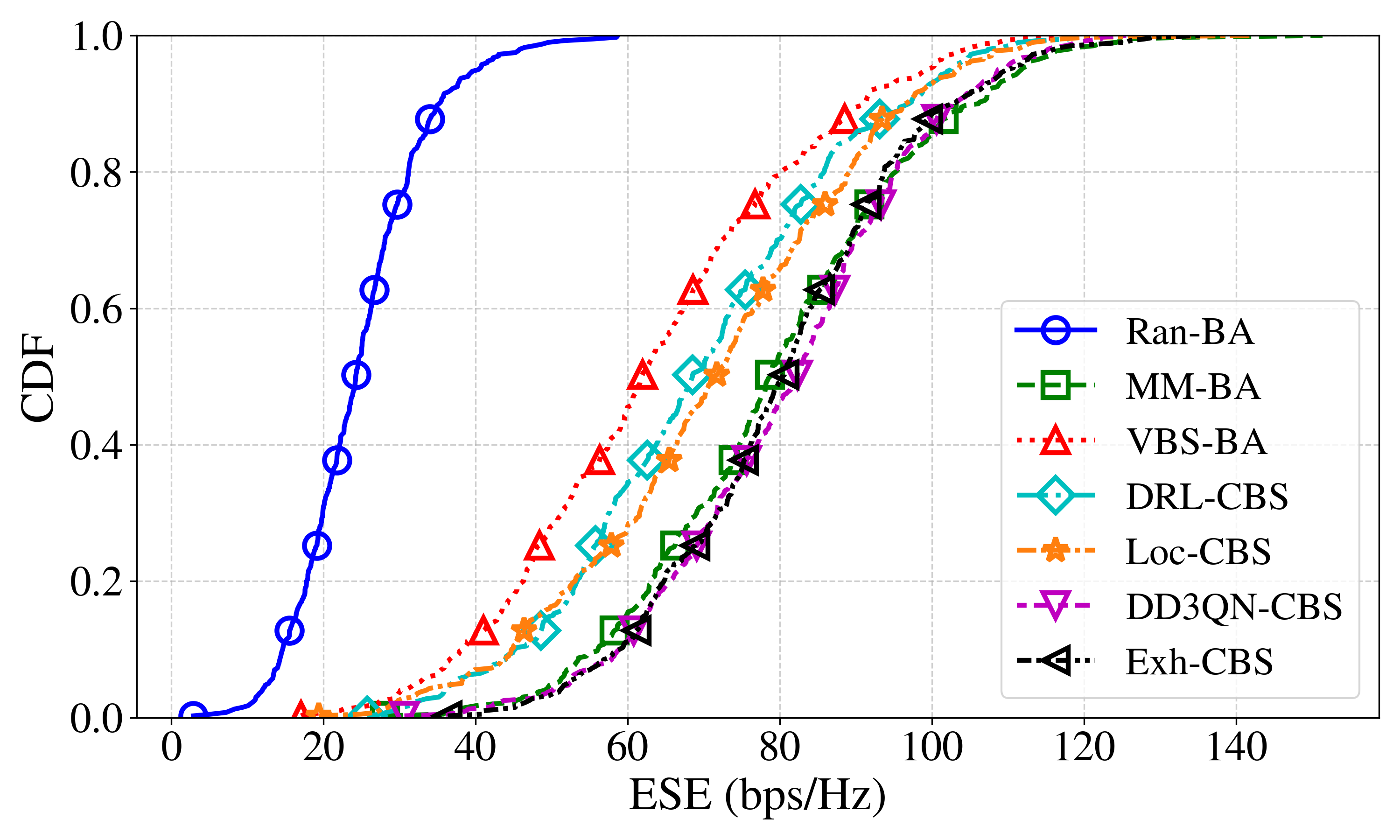}
        \caption{CDF with $N_\mathrm{RF}=14$.}
        \label{fig:test_CDF_14_64_10_-174}
    \end{subfigure}
    \begin{subfigure}[b]{0.48\linewidth}
        \centering
        \includegraphics[width=\linewidth]{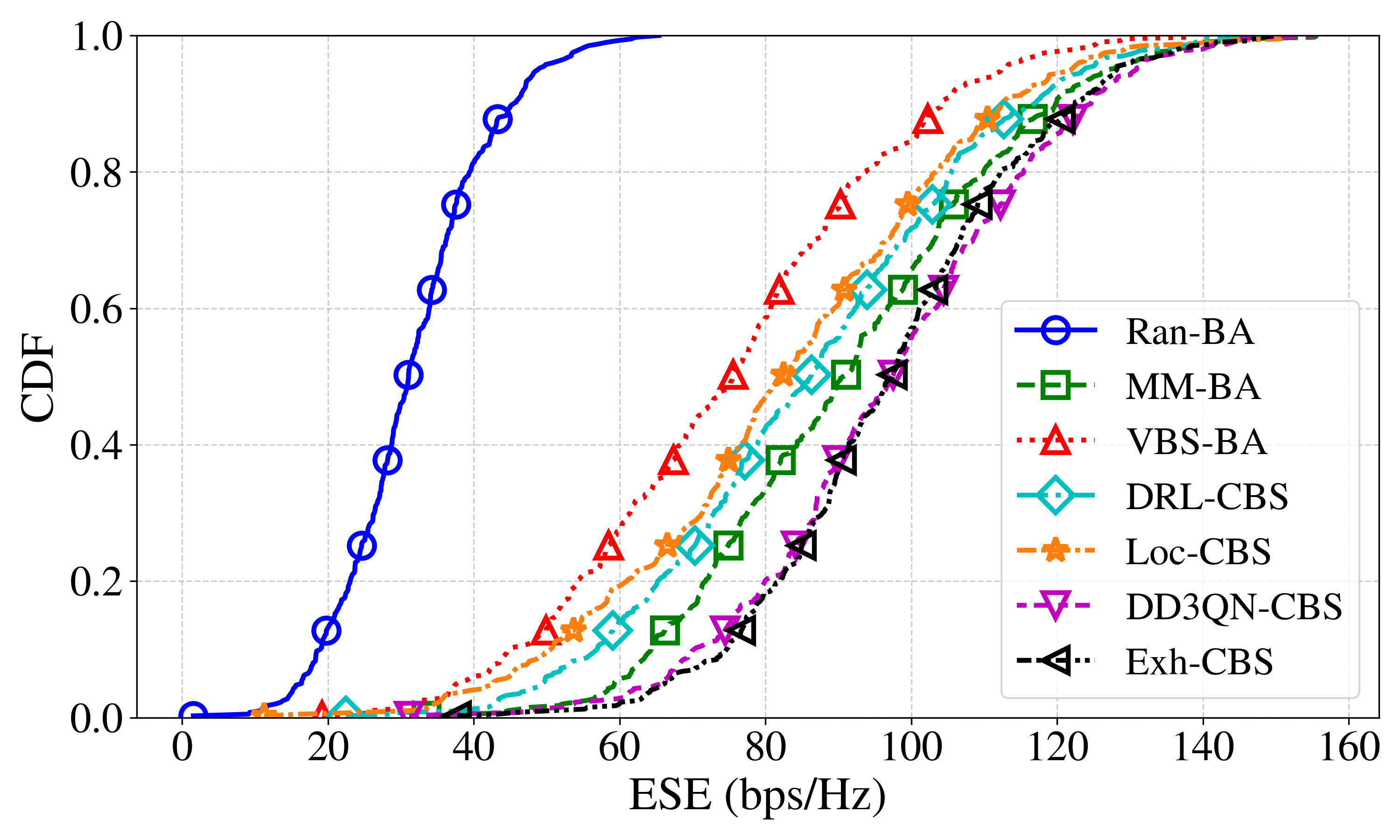}
        \caption{CDF with $N_\mathrm{RF}=18$.}
        \label{fig:test_CDF_18_64_10_-174}
    \end{subfigure}
    \begin{subfigure}[b]{0.48\linewidth}
        \centering
        \includegraphics[width=\linewidth]{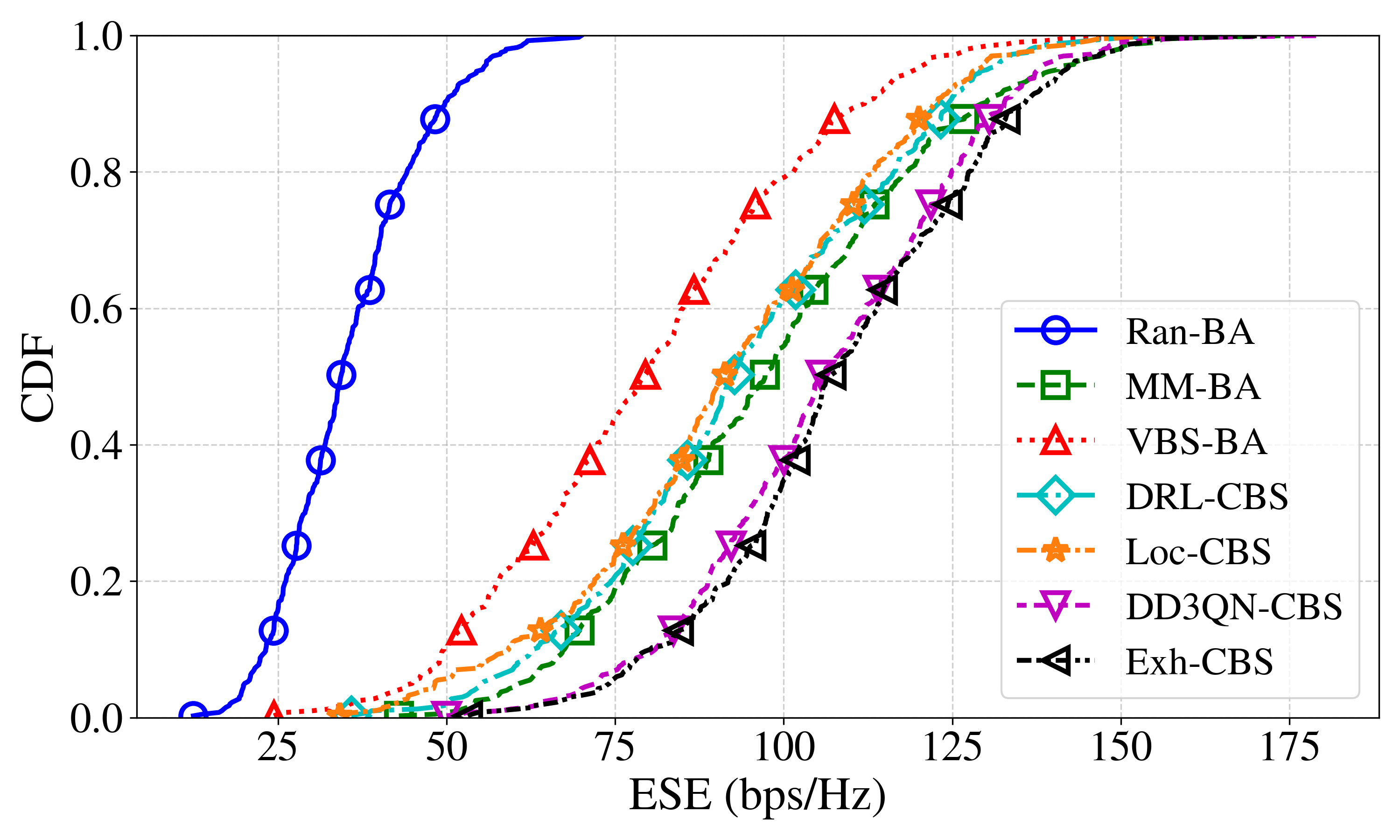}
        \caption{\textcolor{black}{CDF with $N_0=-174$ dBm/Hz.}}
        \label{fig:test_CDF_20_64_10_-174}
    \end{subfigure}
    \begin{subfigure}[b]{0.48\linewidth}
        \centering
        \includegraphics[width=\linewidth]{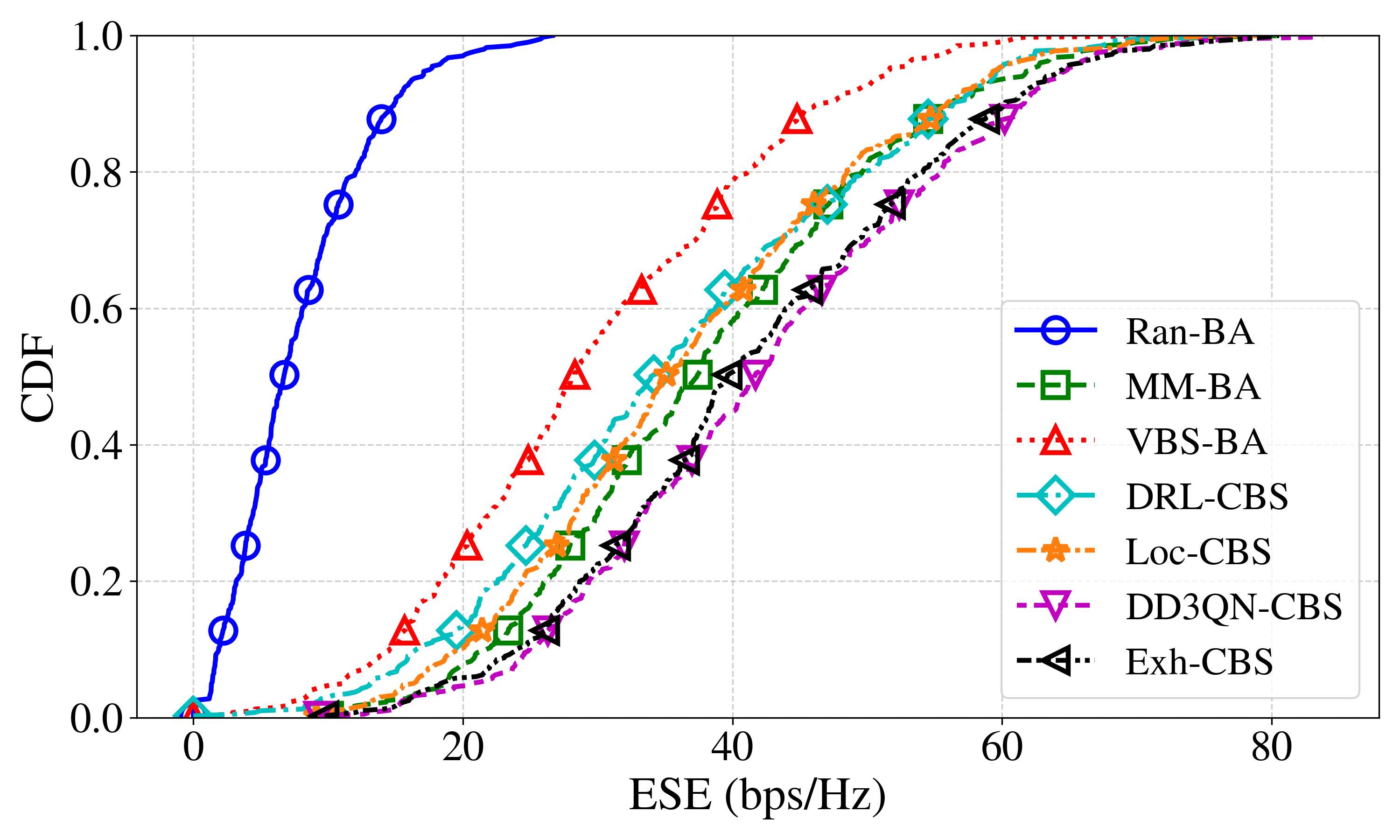}
        \caption{\textcolor{black}{CDF with $N_0=-154$ dBm/Hz.}}
        \label{fig:test_CDF_20_64_10_-154}
    \end{subfigure}
    \begin{subfigure}[b]{0.48\linewidth}
        \centering
        \includegraphics[width=\linewidth]{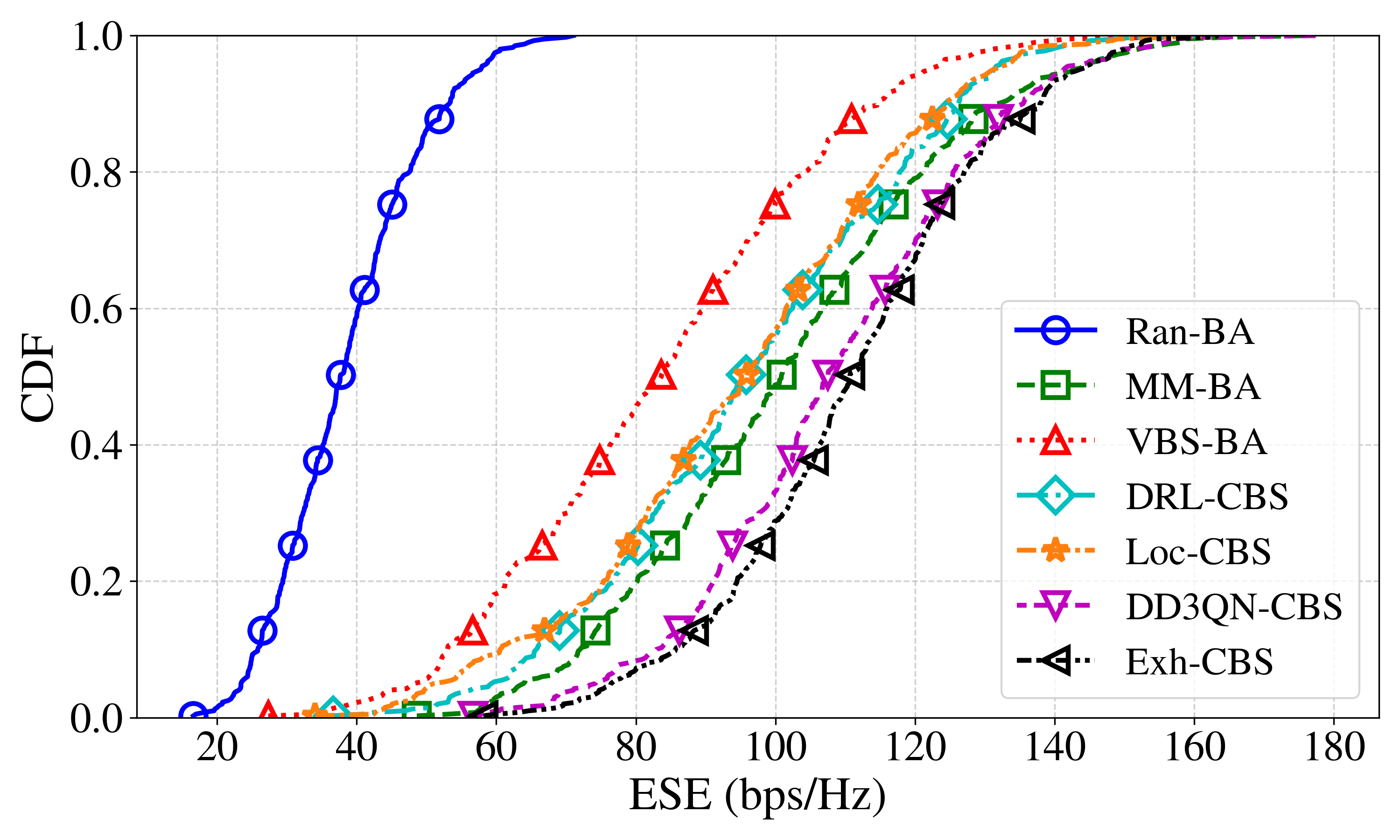}
        \caption{\textcolor{black}{CDF with $\mathrm{SINR}_\mathrm{thres}=5$ dB.}}
        \label{fig:test_CDF_20_64_5_-174}
    \end{subfigure}
    \begin{subfigure}[b]{0.48\linewidth}
        \centering
        \includegraphics[width=\linewidth]{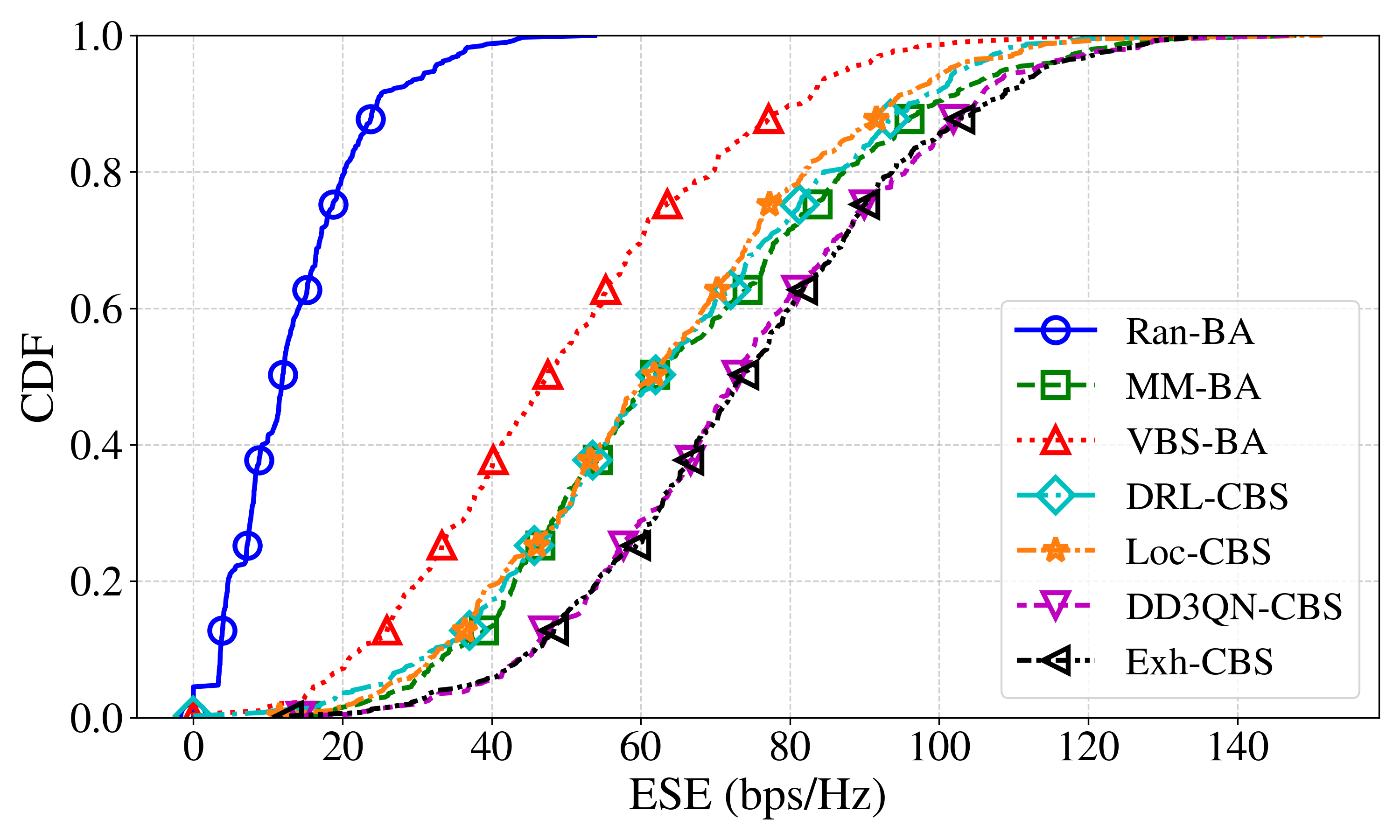}
        \caption{\textcolor{black}{CDF with $\mathrm{SINR}_\mathrm{thres}=30$ dB.}}
        \label{fig:test_CDF_20_64_30_-174}
    \end{subfigure}
    \caption{\textcolor{black}{CDFs of test ESE under different system configurations.}}
    \label{fig:CDF_beam_test_comparison}
\end{figure}

DD3QN-CBS is benchmarked against stochastic, heuristic, and learning-based baselines. 1) \textbf{Ran-BA} randomly assigns BS and user beam pairs. 2) \textbf{VBS-BA} applies the maximum-magnitude (MM) criterion \cite{mm_selection} to a fused magnitude representation, where noisy measured magnitudes are used at the initially scanned entries and VBS-predicted magnitudes are retained elsewhere; 3) \textbf{MM-BA} applies the same MM criterion to the beamspace with full measurements; 4) \textbf{DRL-CBS} serves as a
learning-based baseline using the D3QN architecture \cite{joint_beam_selection_precoding} for
beam selection. Since this algorithm was originally designed
for single-antenna users with joint precoding, we adapt it
by selecting the user-side beam with the largest virtual
gain under MMSE precoding. 5) \textcolor{black}{\textbf{Loc-CBS} employs the same hierarchical DD3QN architecture as the proposed scheme while constructing its beamspace features from BS and user locations. Loc-CBS thus provides a learning-based location-only comparison; 6) \textbf{Exh-CBS} obtains the complete measured beamspace through exhaustive $128\times8$ beam scanning and then performs coordinated beam selection without a VBS-derived beamspace prior.} We evaluate the schemes under different numbers of RF chains $N_\mathrm{RF}$, noise power densities $N_0$, and QoS thresholds $\mathrm{SINR}_\mathrm{thres}$. Unless otherwise specified, the default configuration is $N_\mathrm{RF}=K=20$, $N_0=-174$ dBm/Hz, and $\mathrm{SINR}_\mathrm{thres}=10$ dB.

\subsubsection{Convergence Performance}

\textcolor{black}{As illustrated in Fig.~\ref{beam_training_ese_comparison},
DD3QN-CBS exhibits more sustained improvements in training
ESE across the evaluated configurations. While DRL-CBS
achieves comparable ESE during the early training episodes,
its subsequent improvements gradually diminish, whereas
DD3QN-CBS continues to improve and achieves higher training
ESE in the later stages. The leader-follower decision process
and action masking restrict subsequent selections to valid
beam assignments, allowing the agents to focus on feasible
beam combinations. This design is consistent with the more
sustained training improvement observed for DD3QN-CBS.}

\subsubsection{Average ESE under Different Configurations}
\textcolor{black}{Fig.~\ref{fig:average_ESE_comparison} shows the average ESE under different system configurations. DD3QN-CBS achieves higher
average ESE than Ran-BA, MM-BA, VBS-BA, DRL-CBS, and
Loc-CBS across all evaluated conditions. Among the
magnitude-based baselines, MM-BA consistently outperforms
VBS-BA, benefiting from full beamspace measurements rather
than the combination of partial measurements and VBS-derived
estimates. The further improvement achieved by DD3QN-CBS
is consistent with the benefit of coordinating beam assignments
under inter-user interference instead of relying solely on
individual beam magnitudes.}

\textcolor{black}{Fig.~\ref{fig:N_rf_average_ESE} evaluates the effect of
$N_\mathrm{RF}=K$ on the average ESE. Increasing the number
of RF chains allows more users to be served simultaneously,
leading to higher aggregate ESE. The performance gap between
DD3QN-CBS and MM-BA widens as $K$ increases, reflecting
the growing importance of coordinated beam assignments when
more users share the available spatial resources. Exh-CBS
achieves higher ESE in several configurations, particularly
with larger numbers of scheduled users, while DD3QN-CBS
remains competitive with partial beam measurements.}

\textcolor{black}{Fig.~\ref{fig:Pnoise_average_ESE} shows the effect of noise
power spectral density $N_0$. As the noise power increases,
fewer users satisfy the SINR requirement, reducing the ESE
of all schemes. The performance gap between DD3QN-CBS
and MM-BA gradually narrows under stronger noise, as fewer
beam assignments can support links meeting the QoS threshold.
Exh-CBS achieves higher ESE at the lowest evaluated noise
power spectral density, whereas DD3QN-CBS achieves comparable 
ESE at the other tested noise conditions.}

\textcolor{black}{Fig.~\ref{fig:SINR_average_ESE} examines the effect of the
QoS threshold $\mathrm{SINR}_\mathrm{thres}$. Increasing the
threshold excludes more users from contributing to the ESE,
particularly those experiencing weak desired signals or strong
inter-user interference. The advantage of DD3QN-CBS over
MM-BA increases over the lower and intermediate thresholds
before narrowing under more stringent QoS requirements.
Exh-CBS achieves higher ESE at some evaluated thresholds,
including several lower thresholds, while DD3QN-CBS
performs competitively at others.}

\textcolor{black}{The comparison with Loc-CBS evaluates the benefit of
VBS-derived propagation information beyond location-based
learning. At the default setting, DD3QN-CBS achieves an
average ESE of 106.621 bps/Hz, compared with 91.949 bps/Hz
for Loc-CBS, corresponding to a 15.96\% improvement.
DD3QN-CBS also achieves higher average ESE across all
18 evaluated operating conditions, indicating that the
reflection geometry represented by the VBS database provides
useful information for coordinated beam selection.}

\textcolor{black}{We further examine VBS-guided candidate pruning by comparing
DD3QN-CBS with $64\times4$ and $128\times8$ candidate spaces.
Both configurations use the same VBS-derived beamspace prior
and partial beam-training budget. The $64\times4$ configuration
contains 75\% fewer candidate beam pairs. At the default
setting, the two configurations achieve average ESE values
of 106.621 and 105.794 bps/Hz, respectively, showing that
the reduced candidate space preserves competitive beam
selection performance.}

\textcolor{black}{Exh-CBS performs exhaustive $128\times8$ beam scanning
without VBS assistance and achieves 108.556 bps/Hz at the
default setting. Although complete beamspace measurements
provide higher ESE in some conditions, the proposed
DD3QN-CBS achieves comparable or higher ESE in the
remaining conditions with a substantially smaller candidate
space and fewer online beam measurements. This comparison
illustrates the tradeoff between beam selection performance
and training overhead.}

\subsubsection{Distribution of ESE under Different Configurations}

\textcolor{black}{Fig.~\ref{fig:CDF_beam_test_comparison} presents the cumulative
distribution functions (CDFs) of the test ESE samples under
different system configurations. Compared with Ran-BA,
MM-BA, VBS-BA, and DRL-CBS, DD3QN-CBS generally shifts
the ESE distributions toward higher values. The improvement
also extends to the lower-ESE region, indicating that the
performance gain is not confined to samples with favorable
beam assignments.}

\textcolor{black}{DRL-CBS employs a penalty-based reward under D3QN architecture to discourage beam
collisions and handle inter-user interference, which can place greater emphasis on avoiding
conflicting beam assignments than on distinguishing their
interference-limited performance. In contrast, DD3QN-CBS
uses intermediate SINR-aware rewards to guide sequential
beam decisions before the terminal ESE becomes available.
The more favorable ESE distributions are consistent with
the benefit of incorporating interference-aware feedback
into coordinated beam selection.}

\textcolor{black}{Compared with Loc-CBS, the DD3QN-CBS distributions are
consistently shifted toward higher ESE across the displayed
configurations. Since both methods adopt the same
hierarchical DD3QN architecture, this comparison highlights
the additional propagation information provided by the
VBS-derived reflection geometry. The distributions of
DD3QN-CBS and Exh-CBS are much closer, with their relative
ordering varying across operating conditions. While
exhaustive scanning can achieve higher ESE in some cases,
VBS-guided partial beam training maintains competitive
performance with fewer online beam measurements.}

\subsubsection{Computational Complexity and Runtime}

\textcolor{black}{We further analyze the computational complexity of DD3QN-CBS
with respect to the candidate beam dimensions and the
number of scheduled users. Let $B=\tilde{N}_{\mathrm{BS}}$
and $U=\tilde{N}_{\mathrm{UE}}$ denote the numbers of
BS-side and user-side beam candidates, respectively,
while $C$, $F$, and $D$ denote the input channels,
convolutional channels, and hidden dimension of the
Q-networks. Since the hierarchical decision process
performs one BS-side and one user-side selection for
each scheduled user, the overall inference complexity
is $\mathcal{O}(BUK^2F(C+F)+K(B+U)D^2)$.
For fixed candidate dimensions and network widths,
the complexity scales as $\mathcal{O}(K^2)$ with the
number of users. The SE module introduces additional
channel pooling and excitation operations without
changing the leading-order complexity.}

\textcolor{black}{To complement the complexity analysis, we evaluate
the computational runtime of DD3QN-CBS using an
NVIDIA RTX 5880 Ada Generation GPU. For the implemented network with $C=6$, $F=16$,
and $D=128$, each online Q-network contains 104,662
trainable parameters. 
For $K=14$ and $64\times4$ beam candidates,
the average processing time is 39.292 ms,
compared to the 20.113 ms required for the DRL-CBS baseline. This additional computational cost supports
interference-aware coordination across transmit and
receive beam selections, while the resulting beam
configurations achieve higher ESE under the evaluated
conditions.}

\section{Conclusion}
\label{conclusion}

\textcolor{black}{In this paper, we proposed a VBS-guided beam management framework that leverages multi-modal environmental data to reduce online beam training overhead and support coordinated MU-MIMO beam selection. By modeling dominant single-bounce reflections through mirror symmetry across building facades reconstructed from 3D LiDAR point clouds, the framework constructs a compact VBS database that retains representative virtual-source locations and their associated LoS/VLoS coverage regions. The database supports coarse channel reconstruction to generate a VBS-derived beamspace prior, which guides VOP-based partial beam training. DD3QN-CBS then combines this prior with online measurements to coordinate beam selection under inter-user interference. Simulation results demonstrate improved beam candidate coverage and ESE relative to the corresponding location-only and conventional beam-selection baselines. Compared with exhaustive scanning, the proposed framework achieves competitive ESE with fewer online beam measurements. The robustness evaluations further show that VBS-guided candidate selection retains useful beam directions under the examined reflection-loss variations and sensing--channel mismatches. Future work will consider dynamic environments with moving blockers, online VBS database updates, and reflection calibration to improve the fidelity of VBS-derived beamspace priors.}

\normalem
\let\oldbibliography\bibliography
\renewcommand{\bibliography}[1]{%
  \sloppy
  \let\oldstditem\item
  \renewcommand{\item}{\vspace{-0.4ex}\oldstditem}%
  \oldbibliography{#1}%
}
\bibliographystyle{IEEEtran}
\bibliography{IEEEabrv,reference}

\newpage
\clearpage

\end{document}